%% file: arxive_revised.tex
\documentclass[journal,twoside,web]{ieeecolor}
\usepackage{generic,cite,amsmath,amssymb,amsfonts,multicol,tikz,multirow}
\usepackage{algorithmic,graphicx,textcomp, caption, mathtools,booktabs}

\usepackage[font=small]{caption}
\newtheorem{definition}{Definition}
\newtheorem{theorem}{Theorem}
\newtheorem{lemma}{Lemma}
\newtheorem{assumption}{Assumption}
\newtheorem{remark}{Remark}

\DeclareMathOperator*{\minimize}{minimize}
\DeclareMathOperator*{\subject_to}{s. t. }
\def\BibTeX{{\rm B\kern-.05em{\sc i\kern-.025em b}\kern-.08em
    T\kern-.1667em\lower.7ex\hbox{E}\kern-.125emX}}
\usepackage{fancyhdr}
\usepackage{xcolor}

\fancyfootoffset{-1em}
\fancypagestyle{copyright}{
	\chead{\footnotesize \textcopyright 2024 IEEE.  Personal use of this material is permitted. Permission from IEEE must be obtained for all other uses, in any current or future media, including reprinting/republishing this material for advertising or promotional purposes, creating new collective works, for resale or redistribution to servers or lists, or reuse of any copyrighted component of this work in other works.}
}

\begin{document}
	
\title{Robust Data-Driven Moving Horizon Estimation for Linear Discrete-Time Systems}
\author{Tobias M. Wolff, Victor G. Lopez, \emph{Member, IEEE}, and Matthias A. Müller, \emph{Senior Member, IEEE}
\thanks{This project has received funding from the European Research Council (ERC) under the European Union’s Horizon 2020 research and innovation programme (grant agreement No 948679). }
\thanks{Tobias M. Wolff, Victor G. Lopez, and Matthias M. Müller are with the Leibniz University Hannover, Institute of Automatic Control, 30167 Hannover, Germany (email: \{wolff,lopez,mueller\}@irt.uni-hannover.de)}
\thanks{}}

\maketitle

 \thispagestyle{copyright}

\begin{abstract}
In this paper, a robust data-driven moving horizon estimation (MHE) scheme for linear time-invariant discrete-time systems is introduced. The scheme solely relies on offline collected data without employing any system identification step. We prove practical robust exponential stability for the setting where both the online measurements and the offline collected data are corrupted by non-vanishing and bounded noise. The behavior of the novel robust data-driven MHE scheme is illustrated by means of simulation examples and compared to a standard model-based MHE scheme, where the model is identified using the same offline data as for the data-driven MHE scheme.
\end{abstract}


\section{Introduction}
\label{sec:introduction}
\IEEEPARstart{S}{tate} estimation is crucial for many applications such as control, monitoring, and fault diagnosis. For linear systems, the Kalman filter \cite{Kalman1960} and the Luenberger observer \cite{Luenberger1971} are the most commonly applied state estimation methods. An alternative is moving horizon estimation (MHE) \cite{Rawlings2020,Rao2001,Alessandri2014,Gharbi2019,Farina2010}, where the state estimation problem is centered around an optimization problem. Loosely speaking, at each time step, an optimization problem is solved to determine a state sequence that satisfies the mathematical model of the system dynamics, some known state constraints and/or disturbance bounds. Moreover, the computed state sequence is optimal with respect to a cost function considering the past output measurements that are inside a moving time window. 
Hence, the standard MHE (as well as the Kalman filter and the Luenberger observer) requires knowledge of a mathematical model describing the underlying system dynamics. 

A framework for an alternative system representation that does not require any system identification was established in \cite{Willems2005}. The main result of that work states that any system trajectory of length $L$ of a linear time-invariant discrete-time (LTI-DT) system can be represented by using the span of a single persistently exciting trajectory measured from that system. This result, also known as Willems' fundamental lemma, has been used for data-driven simulation and control \cite{Markovsky2008}, for the design of data-driven linear quadratic regulators \cite{De2019}, and, among many other applications and extensions, for the design of model predictive control (MPC) schemes \cite{Coulson2019} (with robustness guarantees\cite{Berberich2020_MPC}) purely based on input/output and possibly state data of the system. Note that all of these controller design methods do not employ any system identification step. Further work is summarized in the recent survey concerning data-driven system analysis and control in \cite{Markovsky2021}. All of these aforementioned results are for discrete-time systems. Recently, a continuous-time version of the fundamental lemma has been proposed in \cite{Lopez2022}.

With respect to the dual problem of data-driven state estimation, there are only few results available. One focus in this area is the design of data-driven unknown-input observers \cite{Turan2021,Shi2022}. A data-driven state estimation method that exploits the duality principle of control and estimation is proposed in \cite{Adachi2021}. An approach on how to perform data-driven set-based state estimation (without explicitly employing the fundamental lemma) is presented in \cite{Alanwar2021}. Recently, while this paper was under review, a data-driven version of the Kalman filter was proposed in \cite{liu2023learning}. However, the developed filter does not allow to consider state constraints and the theoretical results rely on the assumption of observability, instead of the less restrictive property of detectability.

The contributions of this paper are the following. We introduce a robust data-driven MHE framework for detectable linear systems and prove practical robust exponential stability (pRES) for the setting where, in addition to the measurement noise affecting the online output measurements, the offline measured state and output sequences are also subject to some bounded, non-vanishing measurement noise.

Note that a preliminary version of a nominal case (i.e., noise-free offline data) is available in the conference proceedings \cite{Wolff2021}. Here, we additionally treat the case of noisy offline data and reduce (some of) the conservatism in the state estimation error bounds. Furthermore, our numerical example is substantially more exhaustive, since, e.g., we here compare the performance of the data-driven MHE scheme to the performance of a model-based MHE scheme.

The outline of this paper is the following. In Section~\ref{sec:preliminaries}, we present the setup of this paper and some technical definitions. In Section~\ref{sec:pRGES_state_output}, we prove pRES of the proposed MHE framework. We close this paper by means of an illustrative example and a conclusion in Sections~\ref{sec:ill_example} and~\ref{sec:Conclusion}, respectively. 


\section{PRELIMINARIES \& SETUP}
\label{sec:preliminaries}
The set of integers in the interval $[a,b] \subset \mathbb{R}$ is denoted by $\mathbb{I}_{[a,b]}$ and the set of integers greater than or equal to $a$ by $\mathbb{I}_{\geq a}$. For a vector $x = [x_1 \dots x_n]^\top \in \mathbb{R}^n$ and a symmetric positive definite matrix $P$, we write $|x|_P = \sqrt{x^\top P x} $. The minimal eigenvalue of $P$ is denoted by $\lambda_{\min}(P)$. The identity matrix of dimension $n$ is denoted by $I_n$. The Euclidean norm $||x||_2$ is written as $|x|$ and the infinity norm $||x||_\infty$ as $|x|_\infty$. A stacked window of a sequence $\{x(k) \}_{k=0}^{N-1}$ is written as $x_{[0,N-1]} = \begin{bmatrix} x(0)^\top & \dots & x(N-1)^\top  \end{bmatrix}^\top$. A function $\gamma : \mathbb{R}_{\geq 0} \rightarrow \mathbb{R}_{\geq 0} $ is of class $\mathcal{K}$, if $\gamma$ is continuous, strictly increasing, and $\gamma(0) = 0$. The maximum of two scalars $a,b \in \mathbb{R}$ is denoted by $\max\{a,b\}$. Furthermore, the Hankel matrix of depth $L$ of a stacked window ${x}_{[0,N-1]}$ is defined by
\begin{equation*}
	H_L({x}_{[0,N-1]}) = \begin{pmatrix}
		x(0) & x(1) & \dots & x(N-L) \\
		x(1) & x(2) & \dots & x(N-L+1)\\
		\vdots & \vdots & \ddots & \vdots \\
		x(L-1) & x(L) & \dots & x(N-1) \\
	\end{pmatrix} \hspace{-4pt}.
\end{equation*}

All of our results are crucially based on Willems' fundamental lemma \cite{Willems2005}. It holds for LTI-DT systems of the form
\begin{subequations} 
	\begin{align}
		x(t+1) &= Ax(t) +Bu(t)\\
		y(t) &= Cx(t) + Du(t),
	\end{align}
	\label{system}
\end{subequations}
where $u(t) \in \mathbb{R}^m$, $x(t) \in \mathbb{R}^n$, and $y(t) \in \mathbb{R}^p$. 
The fundamental lemma exploits the notion of persistency of excitation.
\begin{definition}
	An input sequence $\{u(k)\}_{k=0}^{N-1}$ is persistently exciting of order $L$ if $\mathrm{rank}(H_L(u_{[0,N-1]})) = mL$.
\end{definition}

Besides the fundamental lemma, the authors in \cite{Willems2005} prove the following lemma.
\begin{lemma}
	\label{Lemma_Rank}
	Suppose $\{x(k)\}_{k=0}^{N-1}$, $\{u(k)\}_{k=0}^{N-1}$ is an input/state trajectory of the controllable LTI system (\ref{system}) where $\{u(k)\}_{k=0}^{N-1}$ is persistently exciting of order $L + n$, then the matrix
	\begin{equation}
		\begin{bmatrix}
			H_1(x_{[0,N-L]}) \\
			H_L(u_{[0, N-1]})
		\end{bmatrix}
	\end{equation}
	has full row rank.
\end{lemma}

The fundamental lemma can be formulated in the classical state space framework as follows \cite{Berberich2020_trajectory,VanWarde2020}.
\begin{theorem}
	(\hspace{1sp}\cite{Berberich2020_trajectory}) Suppose $\{u(k)\}_{k=0}^{N-1}$, $\{y(k)\}_{k=0}^{N-1}$ is an input/output trajectory of a controllable LTI system (\ref{system}), where $\{u(k)\}_{k=0}^{N-1}$ is persistently exciting of order $L+n$. Then, $\{\overline{u}(k)\}_{k=0}^{L-1}$, $\{\overline{y}(k)\}_{k=0}^{L-1}$ is a trajectory of system (\ref{system}) if and only if there exists $\alpha \in \mathbb{R}^{N-L +1}$ such that
	\begin{equation}
		\begin{bmatrix}
			H_L(u_{[0,N-1]}) \\
			H_L(y_{[0,N-1]}) \\
		\end{bmatrix} \alpha = \begin{bmatrix}
			\overline{u}_{[0,L-1]}\\
			\overline{y}_{[0,L-1]} \\
		\end{bmatrix} \label{Willems_Lemma} .
	\end{equation}
	\vspace{0.01pt}
\end{theorem}

This important result states that any length $L$ system trajectory of a linear controllable LTI system can be expressed in terms of a single persistently exciting system trajectory. This implies that one does not necessarily need a mathematical model of the treated system to represent system trajectories. Particularly relevant in the context of state estimation is the result summarized in the following remark.

\begin{remark}
	\label{Minimal_realization}
	In addition to (\ref{Willems_Lemma}), it holds that \cite[Eq. (5)]{Berberich2020_trajectory}
	\begin{equation}
		\overline{x}_{[0, L-1]} = H_L(x_{[0, N-1]}) \alpha,
	\end{equation}
	where $\overline{x}_{[0,L-1]}$ is the state trajectory of system (\ref{system}) corresponding to $\overline{u}_{[0,L-1]}$, $\overline{y}_{[0,L-1]}$ of (\ref{Willems_Lemma}) and $x_{[0,N-1]}$ the state trajectory corresponding to $u_{[0,N-1]}$, $y_{[0,N-1]}$.
\end{remark}

Throughout this paper, we consider an offline phase and an online phase. In the offline phase, we collect (noise-free) input, (noisy) output and (noisy) state trajectories of length $N$ from the considered system so that an implicit system representation according to the fundamental lemma can be established. That is, we measure the noisy a priori state (denoted by $\tilde{x}^d$) and output (denoted by $\tilde{y}^d$) data 
	\begin{align}
		\tilde{y}^d(t) &= y^d(t) + \varepsilon_y^d(t) \label{offline_outputs}\\ 
		\tilde{x}^d(t) &= x^d(t) + \varepsilon_x^d(t) \label{offline_states},
	\end{align}
	where $\varepsilon_x^d$ and $\varepsilon_y^d$ are the offline state and output measurement noise with $|\varepsilon_x^d(t)|_\infty \leq \overline{ \varepsilon}_x^d$ and $|\varepsilon_y^d(t)|_\infty \leq \overline{\varepsilon}_y^d$, respectively. The noise-free outputs and states are denoted by $y^d$ and $x^d$, respectively. Furthermore, we define
	\begin{equation}
		\overline{ \varepsilon}^d \coloneqq \max \{\overline{\varepsilon}_y^d, \overline{\varepsilon}_x^d \}. \label{def:eps_bar}
\end{equation}

In the online phase, the states are no longer measurable and need to be estimated from input and output measurements, where the outputs are corrupted by some non-vanishing measurement noise $v$. Therefore, the system dynamics considered in the online phase are
\begin{subequations} \label{system_def}
	\begin{align}
		x(t+1) &= Ax(t) +Bu(t) \label{state_dynamics}\\
		\tilde{y}(t) &= Cx(t) + Du(t) + v(t)  
	\end{align}
\end{subequations}
that differs from (\ref{system}) by the non-vanishing measurement noise $v(t) \in \mathbb{V} \subset \mathbb{R}^p$.

The assumption that noisy state measurements are available in an offline phase could be restrictive in general. However, it is satisfied in a broad range of applications where the states are measurable using, e.g., a dedicated laboratory and specific hardware, which is not available in online operations of the system (compare also the discussion in, e.g., \cite{Turan2021}). For instance, this assumption is fulfilled in the application of autonomous driving, where the states of a vehicle can be measured by expensive sensors in an offline phase, compare \cite{Graeber2019,Ehlers2022}. Due to the high costs, these sensors are not placed in the vehicles in a series production. Hence, in the online phase (i.e., when the expensive sensors are no longer available), the states must be estimated. This assumption is indispensable for the results of our work. Loosely speaking, the online input and output measurements are combined with the offline collected input and output measurements to compute a vector~$\alpha$. This~$\alpha$, together with the offline collected state trajectory, generates the estimated state trajectory. This estimated state trajectory is in the same (unknown) realization~(\ref{system}) as the offline collected state trajectory, compare Remark~\ref{Minimal_realization}. Note that without such an offline collected state trajectory, the (online) state trajectory could not be correctly estimated since the realization~($A$,$B$,$C$,$D$) of system (\ref{system}) is unknown (and hence different state trajectories can explain the given input/output data), as is, e.g., well known from the context of subspace identification \cite{Van1997}. 

\section{PRACTICAL RES FOR NOISY OFFLINE STATE \& OUTPUT MEASUREMENTS}
\label{sec:pRGES_state_output}
In this section, we introduce the robust data-driven MHE scheme.  Given an estimation horizon $L$ in $\mathbb{I}_{> 0 }$, at each time\footnote{For $t<L$, replace each $L$ in (\ref{MHE_noisy}) by $t$, i.e., take all available measurements into account.} $t \geq L$, solve
	\begin{subequations} \label{MHE_noisy}
	\begin{align}
		&\minimize_{\substack{\overline{x}_{[-L,0]}(t), \alpha(t), \\ \sigma_{[-L,-1]}^y (t),\\ \sigma_{[-L,0]}^x (t)}} \hspace{0.1cm}  J(\overline{x}(t-L|t),\sigma^y_{[-L, -1]}(t),\sigma^x_{[-L, 0]}(t), \alpha(t)) \label{cost_function_noisy} \\
		&\subject_to \hspace{0.2cm} \begin{bmatrix}
			H_L(u^d_{[0,N-2]}) \\
			H_L(\tilde{y}^d_{[0,N-2]}) \\
			H_{L+1}(\tilde{x}^d_{[0,N-1]}) \\
		\end{bmatrix}
		\alpha(t) =
		\begin{bmatrix}
			u_{[t-L,t-1]}\\
			\tilde{y}_{[t-L,t-1]} -\sigma_{[-L,-1]}^y (t)\\
			\overline{x}_{[-L ,0]}(t) + \sigma_{[-L,0]}^x (t)\\
		\end{bmatrix} \label{System_Dynamics_noisy}\\
		& \hspace{2cm} \overline{x}_{[-L,0]}(t) \in \mathbb{X} \label{state_constraint_noisy}
	\end{align}
	where
	\begin{align}
		J(\overline{x}(t-L|t),&\sigma^y_{[-L, -1]}(t), \sigma^x_{[-L, 0]}(t), \alpha(t)) \nonumber \\ 
		&\coloneqq  \Gamma(\overline{x}(t-L|t))+ \sum_{k=1}^{L}\ell_k (\sigma^y(t-k|t)) \nonumber \\ 
		&+ c_{\sigma^x}|\sigma^x_{[-L, 0]}(t)|^2+ c_\alpha( \overline{\varepsilon}_x^d + \overline{\varepsilon}_y^d)|\alpha(t)|^2.		
		\label{MHE_noisy_cost}
	\end{align}
\end{subequations}
We denote the online measured inputs and outputs from time~$t-L$ up to time $t-1$ by $u_{[t-L,t-1]} $ and $\tilde{y}_{[t-L,t-1]}$, respectively. The estimated state sequence from time $t-L$ up to time $t$, estimated at time~$t$, is denoted by $\overline{x}_{[-L,0]}(t) \coloneqq [\overline{x}(t-L|t)^\top, \dots, \overline{x}(t|t)^\top ]^\top$. Similarly, the estimated fitting error from time $t-L$ up to time~$t-1$, estimated at time~$t$, is written as $ \sigma_{[-L,-1]}^y(t)$. We introduce the fitting error $\sigma^y_{[-L,-1]}(t)$ in the optimization problem since the noisy online output measurements $\tilde{y}$ might not be in the span of the offline collected outputs $y^d$ (similar to data-driven MPC, where the future predicted outputs might not be in the span of the offline collected noisy data \cite{Coulson2019,Berberich2020_MPC}). Since the offline collected state data is also noisy, we consider a slack variable $\sigma^x$ in the last block row of (\ref{System_Dynamics_noisy}). Without $\sigma^x$, we cannot necessarily guarantee that the estimated state sequence respects the constraints (\ref{state_constraint_noisy}). The left-hand side of (\ref{System_Dynamics_noisy}) contains the Hankel matrices (of depth $L$ or $L+1$) of the offline collected inputs, outputs, and states. The variable $u^d$ stands for the offline collected input data and $\mathbb{X}$ for the state constraint set. As usual in MHE \cite{Mueller2017,Allan2019}, we can incorporate knowledge about the estimated state sequence in form of state constraints in order to improve the estimation performance, compare (\ref{state_constraint_noisy}). These constraints are usually inherently satisfied by the system states such as, e.g., nonnegativity constraints of concentrations in a chemical reactor or serum hormone concentrations.  
In the cost function~(\ref{MHE_noisy_cost}), we consider four terms. First, the prior weighting, which is defined as
\begin{equation}
	\label{Prior_Weighting_nom}
	\Gamma (\overline{x}(t - L|t)) \coloneqq |\overline{x}(t-L|t) - \hat{x}(t-L)|_P^2 \rho^{L},
\end{equation}
for a weighting matrix $P$ and the same discount factor $\rho \in (0,1)$. The prior weighting penalizes the difference between the estimated state at the beginning of the horizon, i.e., $\overline{x}(t-L|t)$, and the prior estimate\footnote{This prior is typically called ``filtering prior" in the MHE literature, compare, e.g., \cite{Allan2019}.} $\hat{x}(t-L)$ corresponding to the state estimate at time $t-L$. The prior weighting hence takes into account previous measurements that are not part of the current horizon. Second, we penalize the fitting error $\sigma^y$ in the cost function~(\ref{MHE_noisy_cost}), so that the difference between the online measured outputs and the reconstructed outputs remains small. The stage costs $\ell_k$ are defined as
\begin{equation}
	\ell_k(\sigma^y(t-k|t))	\coloneqq  |\sigma^y(t-k|t)|^2_R \rho^k \label{eq:stage_costs}
\end{equation}
for a weighting matrix $R$ and the same discount factor $\rho$. In the MHE literature, the application of discounting (inducing a fading memory effect) in the disturbance terms has been introduced in \cite{Knufer2018} and has been proven to be particularly useful for various MHE stability results, compare \cite{Knuefer2021,Schiller2022suboptimal}. By means of the matrices $R$ and $P$, we trade off how much we believe our measurements and how much we believe our prior\footnote{For instance, consider the situation, where a good prior estimate and (only) noisy measurements are available, then a suitable choice for the weighting matrices is $R<< P$.}. Third, the cost function considers two quadratic regularization terms to limit the values of $|\sigma^x_{[-L,-1]}(t)|$ and $|\alpha(t)|$.  The regularization related to $\sigma^x$ is incorporated to bring the estimated state trajectory closer to the span of the noisy collected state sequence. The regularization related to $\alpha$ is needed to limit the amplification of the noise affecting the offline collected data. This is similar in spirit to the dual problem of data-driven MPC, where such regularizations are particularly common, compare \cite{Coulson2019,Coulson2019b,Berberich2020_MPC}. To guarantee pRES of the data-driven MHE scheme, the constants $c_\alpha$ and $c_{\sigma^x}$ need to satisfy $ c_{\sigma^x} \geq 2p_0$ and $c_\alpha \geq \max \{\frac{\eta - \eta^{L+1}}{1-\eta}r_0\sqrt{p}(N-1), 2p_0\sqrt{nN} \}$, where the constants $p_0$, $r_0$, and $\eta$ are from Definition~\ref{Definition_UOSS_Text} below.  
The optimizers of problem (\ref{MHE_noisy}) are denoted by $\hat{x}_{[-L,0]}(t)$, $\hat{\sigma}^y_{[-L,-1]}(t)$, $\hat{\sigma}^x_{[-L, 0]}(t)$, $ \hat{\alpha}(t)$ and the related cost by $J^\ast \coloneqq J(\hat{x}_{[-L,0]}(t), \hat{\sigma}^y_{[-L,-1]}(t), \hat{\sigma}^x_{[-L, 0]}(t), \hat{\alpha}(t))$. The state estimate at time $t$ is defined as $\hat{x}(t) \coloneqq \hat{x}(t|t)$, corresponding to the last element of the optimal sequence $\hat{x}_{[-L,0]}(t)$.

In the following, we prove pRES of the data-driven MHE scheme. To simplify the notation, we denote sequences of finite or infinite length by bold face symbols $\mathbf{v} \coloneqq \{ v(t_1), \dots, v(t_2)\}$ for some $t_1$, $t_2 \in \mathbb{I}_{\geq 0}$ or $\mathbf{v} \coloneqq \{ v(t_1), v(t_1 +1 ), \dots\}$. The length of these sequences will be understood from the context. The solution to (\ref{system_def}) at time $t$ for initial condition $x_0$ and input sequence $\mathbf{u}$ is denoted by $x(t; x_0, \mathbf{u})$. We denote the corresponding output as $y(t) = Cx(t)+Du(t) =: h(x(t; x_0, \mathbf{u}), u(t))$. 

In order to prove RGES, we will assume (compare Assumption~\ref{ass_detec} below) that the considered system (\ref{system_def}) is detectable, which implies ``incremental exponential uniform output-to-state stability" (e-UOSS) \cite[Corollary 7]{Knufer2020}, \cite{Rawlings2020} as defined below.
\begin{definition}
	\label{Definition_UOSS_Text}
	The linear system (\ref{system_def}) is e-UOSS if there exist constants  $p_0, r_0 \geq~1$, $\eta \in (0,1)$ such that for each pair of initial conditions $x_1$, $x_2 \in \mathbb{R}^n$ and any input sequence $\mathbf{u} $ generating the states  $x(t; x_1, \mathbf{u})$ and $x(t; x_2, \mathbf{u})$ the following holds for all $t \in \mathbb{I}_{\geq 0}$:
	\begin{align}
		|x(t; &x_1, \mathbf{u}) - x(t; x_2,\mathbf{u})| \leq  p_0|x_1 -x_2| \eta^t + \nonumber \\
		& \sum_{\tau = 1}^{t} r_0|h(x(t- \tau; x_1, \mathbf{u}), u(t-\tau)) \nonumber \\
		& \quad \quad \quad -h(x(t- \tau; x_2, \mathbf{u}), u(t - \tau))| \eta^\tau.  \label{Definition_UOSS}
	\end{align}
\end{definition}
The notion of ``uniformity" is used since the right-hand side of (\ref{Definition_UOSS}) holds uniformly for all $\mathbf{u}$ (with the same $p_0$, $r_0$, and~$\eta$). 

\begin{remark}
	In many model-based MHE stability proofs, the notion of incremental input/output-to-state stability (i-IOSS) is employed. The e-UOSS property considered here differs from that notion by (i) using the exponential version which is always satisfied for linear detectable systems (compare, e.g., \cite[Corollary 7]{Knufer2020}) and by (ii) only considering an output term in (\ref{Definition_UOSS}), but not an additional disturbance input. The extension of our results to this additional disturbance input (also referred to as process noise) is an interesting issue for future research, compare also Remark~\ref{rmk:Process_Noise} in Section \ref{sec:pRGES_state_output}. 
\end{remark}

In order to prove RGES, we need the following four standard assumptions.
\begin{assumption}
	\label{ass_pers}
	The considered system (\ref{system_def}) is controllable and the offline collected input $\{ u^d(k)\} _{k=0}^{N-1}$ is persistently exciting of order $L + n + 1$.
\end{assumption}
Note that this assumption implies $N \geq (m+1)(L+n+1)-1$.
\begin{assumption}
	\label{ass_detec}
	The pair $(A,C)$ of system (\ref{system_def}) is detectable and the discount factor $\rho$ in (\ref{eq:stage_costs}) and (\ref{Prior_Weighting_nom}) is selected such that $\eta \leq \rho$, for $\eta$ in (\ref{Definition_UOSS}).
\end{assumption}
The assumption on the relation of the discount factors is needed for technical reasons that become clear in the proof of Theorem~\ref{thm:pRGES}, see Appendix~\ref{proof}.

Recall the definitions of the stage costs $\ell_k$ in (\ref{eq:stage_costs}) and of the prior weighting $\Gamma$ in (\ref{Prior_Weighting_nom}). Note that the quadratic expressions in the definitions of the stage costs and the prior weighting can always be lower bounded if the matrices $R$ and $P$ are positive definite, i.e., $r_1 |\sigma^y(t-\tau|t)|^2 \rho^k \leq |\sigma^y(t-\tau|t)|_R^2 \rho^k$ and $p_1  |\overline{x}(t-L|t) - \hat{x}(t-L)|^2 \rho^L \leq  |\overline{x}(t-L|t) - \hat{x}(t-L)|_P^2 \rho^L$ with $ r_1 \coloneqq \lambda_{\mathrm{min}}(R)$ and $p_1 \coloneqq  \lambda_{\mathrm{min}}(P)$, respectively.

\begin{assumption}
	\label{ass_stage}
	The matrices $R$ in (\ref{eq:stage_costs}) and $P$ in (\ref{Prior_Weighting_nom}) are positive definite. Furthermore, it holds that $p_0 \leq p_1$ and $r_0 \leq r_1$ with $p_0$ and $r_0$ as from Definition~\ref{Definition_UOSS_Text}. 
\end{assumption}
In order to satisfy Assumptions~\ref{ass_detec} and~\ref{ass_stage}, the constants $r_0$, $p_0$, and $\eta$ from Definition~\ref{Definition_UOSS_Text} need to be known. In Appendix~\ref{sec:db_eUOSS}, we briefly outline how these can be computed in a data-driven fashion.

In order to show pRES of the introduced data-driven MHE~(\ref{MHE_noisy}), we need one further assumption. 
\begin{assumption}
	The inputs $u$ and the states $x$ of the considered system (\ref{system_def}) evolve in compact sets $\mathbb{U}$ and $\tilde{\mathbb{X}}$, i.e, $u(t) \in \mathbb{U}$ and $x(t) \in \tilde{\mathbb{X}}$, respectively.
	\label{ass_compact}
\end{assumption}
Assumption~\ref{ass_compact} implies that the states of system (\ref{system_def}) do not grow unboundedly. If necessary, a pre-stabilizing controller can be applied to system (\ref{system_def}) to ensure that Assumption~\ref{ass_compact} holds. Furthermore, in most practical applications the control actuators have physical limits implying that one can find a compact set $\mathbb{U}$, in which the inputs $u$ evolve.

The main objective of this section is to show that the introduced robust data-driven MHE scheme in (\ref{MHE_noisy}) is pRES with respect to the following definition.
\begin{definition}	\label{def:pRES}
	Consider system (\ref{system_def}) subject to disturbances~$v \in \mathbb{V}$. Moreover, consider noisy a priori output and state measurements as defined in (\ref{offline_outputs}) and (\ref{offline_states}), respectively. A state estimator is pRES if there exist constants $c_1$, $c_2 \geq 1$, $\lambda_1$, $\lambda_{2}\in (0,1)$ and a function $\gamma \in \mathcal{K}$ such that for all $x_0$, $\hat{x}(0) \in \tilde{\mathbb{X}}$, all $v \in \mathbb{V}$, and all $\overline{\varepsilon}_x^d$, $\overline{\varepsilon}_y^d  \geq 0$ the following is satisfied for all $t \in \mathbb{I}_{\geq 0}$:
	\begin{align}
		|x(t) - &\hat{x}(t)| \nonumber \\ \leq  &c_{1} |x_0 - \hat{x}(0)| \lambda_{1}^t + \: \sum_{\tau = 0}^{t} c_{2} |v(t-\tau)| \lambda_{2}^\tau + \gamma( \overline{\varepsilon}^d).  \label{eq:pRGES}
	\end{align}
\end{definition}
Definition~\ref{def:pRES} states that we can upper bound the differences between the true (unknown) states and the estimated states by means of (i) their difference in the initial conditions, (ii) the true measurement noise affecting the online output measurements, and (iii) a $\mathcal{K}$ function depending on $\overline{\varepsilon}^d$ in (\ref{def:eps_bar}). 

Furthermore, this definition states that smaller values for $\overline{\varepsilon}^d$ imply smaller state estimation error bounds. The main result of this paper is given by the following theorem. 
\begin{theorem}
	\label{thm:pRGES}
	(pRES of robust MHE) Consider system (\ref{system_def}) subject to noise $\varepsilon_x^d$ and $\varepsilon_y^d$ in the (offline) data collection phase and to $v\in \mathbb{V}$ in the online phase. Let Assumptions~\ref{ass_pers}~-~\ref{ass_compact} hold. Then, there exists $L_{\min}$ such that for all $L > L_{\min}$ the robust data-driven MHE scheme (\ref{MHE_noisy}) is pRES, i.e., there exist $c_1, c_2 \geq 1$, $\lambda_{1}, \lambda_{2} \in (0,1)$, and a function $\gamma \in \mathcal{K}$, such that for all $x_0 , \hat{x}(0) \in \tilde{\mathbb{X}}$, all $u \in \mathbb{U}$, and for all $t \in \mathbb{I}_{\geq 0}$, (\ref{eq:pRGES}) is satisfied. 
\end{theorem}

The detailed proof can be found in Appendix~\ref{proof}. The idea is to exploit some incremental detectability property (here: e-UOSS) together with bounds on the optimal value function to establish a contraction, which is then applied recursively. The framework of the proof is similar to robust stability proofs of various model-based MHE schemes, see \cite{Knufer2018,Allan2021robust,Mueller2017}. In the here considered data-driven setting with noisy offline data, the following main modifications are necessary compared to standard robust model-based MHE proofs:  
	\begin{itemize}
		\item[-] To bound the optimal value function, a candidate trajectory is needed. The construction of this feasible (but in general suboptimal) candidate trajectory of the MHE problem is here more involved compared to the model-based MHE proofs. We need to choose $\sigma^x$, $\sigma^y$, and~$\alpha$ suitably to define a feasible candidate trajectory that satisfies the state constraints. 
		\item[-] Furthermore, suitable bounds for $|\sigma^x|$, $|\sigma^y|$, and $|\alpha|$ of the candidate solution are needed.
		\item[-] In model-based MHE proofs, one exploits the incremental detectability property by considering the optimal estimated system trajectory and the real (unknown) system trajectory. In the data-driven setting considering noisy offline state and output measurements, the estimated system trajectory might not be a feasible system trajectory anymore. Hence, we need to adapt the estimated trajectory to guarantee that it is a feasible system trajectory such that the incremental detectability property can be exploited. 
		\item[-] Finally, since the offline data is noisy, one can only prove a practical robust stability result that depends on the noise level of the offline data. Noisy offline data are comparable to parametric uncertainty in model-based MHE (see, e.g., \cite{Alessandri2012,Schiller2022moving}), since the Hankel matrices in (\ref{MHE_noisy}) encode an implicit, uncertain ``model" of the system.
\end{itemize}

One inherent drawback of the applied proof technique is that the state estimation bound depends on the \textit{a-priori} upper bound of the offline collected noise, which is more conservative than the actual noise levels $\varepsilon_x^d$ and $\varepsilon_y^d$. This can be explained by the design of the cost function (\ref{MHE_noisy_cost}) which considers the \textit{a-priori} upper bounds of the offline collected noise. Hence, these bounds influence the final state estimation error bounds. 

As is also the case in many model-based MHE stability results, the here applied proof technique in general leads to conservative bounds. Additionally, Theorem~\ref{thm:pRGES} guarantees the existence of a minimal horizon length $L_{\mathrm{min}}$ such that the state estimation error is bounded.  This minimal horizon length $L_{\min}$ can be computed numerically by solving for the smallest $L$ such that (47) in the proof of Theorem~\ref{thm:pRGES} holds. A (more conservative) analytic expression for $L_{\min}$ can also be found, namely by bounding~$\tilde{P}_0$ as defined below (43) with an expression independent of the horizon length $L$.

\begin{remark}
	\label{rmk:nominal_case}
	In case of noise-free offline data, i.e., $\overline{\varepsilon}_x^d = \overline{\varepsilon}_y^d = 0$, we would not consider $\sigma^x$ in the constraints~(\ref{System_Dynamics_noisy}) and in the cost function~(\ref{cost_function_noisy}). Furthermore, no regularization of $\alpha$ is required in the cost function. The final state estimation error bounds are similar, but obviously with $\gamma(0) = 0$ and modified gains, compare \cite{Wolff2021} for details. In this nominal case, the proof technique is closer to the model-based MHE proof techniques, so that none of the adaptions mentioned below Theorem~\ref{thm:pRGES} are necessary.
\end{remark}

\begin{remark}
	\label{rmk:Process_Noise}
	Throughout this work, we do not consider process noise, i.e., noise in the state dynamics of (\ref{system_def}). Process noise is usually considered by an additional term $w(t)$ added to the state equation (\ref{state_dynamics}), resulting in
	\begin{subequations} \label{system_process_noise}
		\begin{align}
			x(t+1) &= Ax(t) +Bu(t) + w(t)\label{system_process}\\
			\tilde{y}(t) &= Cx(t) + Du(t) + v(t) .
		\end{align}
	\end{subequations}
	We here discuss two options how process noise could be treated in the data-driven MHE scheme. First, one could assume that the process noise can be measured directly, as partly assumed in the context of data-driven MPC in, e.g., \cite{Pan2021}. In that case, the process noise could be treated as an additional input, where one would need to make sure that the concatenated control input $u$ and the process noise input $w$ fulfill the persistency of excitation assumption. The constraint~(\ref{state_constraint_noisy}) would need to be augmented by the Hankel matrix containing the offline measured process noise trajectory and (regarding the online phase) by the measured process noise trajectory. However, the assumption that the process noise can be measured in practice is rarely fulfilled. Alternatively, if one does not consider that the process noise can be measured directly, one could incorporate the (unknown) process noise terms in the Hankel matrix composed of the offline measured state trajectory. Namely, if process noise is considered, the entries $\tilde{x}^d(t)$ of the Hankel matrix $H_{L+1}(\tilde{x}^d_{[0,N-1]})$ on the left hand side of (\ref{state_constraint_noisy}) are given by
	\begin{align}
		\tilde{x}^d(t) =& A^t\tilde{x}(0)+\sum_{i=0}^{t-1}A^{t-1-i}Bu^d(i)\nonumber \\
		&+\sum_{i = 0}^{t-1}A^{t-1-i}w^d(i) + \varepsilon_x^d(t).
	\end{align}
	Consequently, the fitting error sequence $\sigma^x$ must compensate (i) the measurement noise $\varepsilon^d_x(t)$ corrupting the offline measured states and (ii) the cumulated process noise. The cumulated process noise would have to be bounded such that the term related to $\sigma^x_{[-L,0]}(t)$ in the cost function remains bounded. Therefore, we would need to upper bound the maximal cumulated process noise at any time $t$ by, e.g.,  
	\begin{align}
		\sum_{i = 0}^{t-1}A^{t-1-i}w^d(i) \leq \sum_{i = 0}^{t-1}|A^{t-1-i}|_\infty w^d_{\max},
	\end{align}
	where $w^d_{\max}$ is the maximal occurring process noise. This is possible if all eigenvalues of the matrix $A$ are strictly inside the unit circle. Therefore, we would need to limit our analysis to stable systems (or pre-stabilized systems). In a related robust stability proof, one would need to treat the cumulated process noise as additional measurement noise and treat it similar to~$\varepsilon^d_x$. 
\end{remark}


\section{APPLICATION TO FOUR-TANK SYSTEM}
\label{sec:ill_example}
In this section, we illustrate the performance of the novel robust data-driven MHE scheme by means of several numerical simulations. We consider exemplarily a four-tank system \cite{Raff2006}. First, we focus on a linearized version of the four-tank system that has also been considered in \cite{Berberich2020_MPC}. Second, we consider the original nonlinear four-tank system. In both cases, we compare the performance of the data-driven MHE scheme to the performance of a model-based MHE scheme where the model is identified by subspace identification (SID) and prediction error minimization (PEM) using the same (offline) input, output and state measurements that are used in the offline phase of the data-driven MHE formulation. In the context of controller design, some recent works compare the performance of direct data-driven control frameworks to indirect data-driven control frameworks, where in the latter the offline data are used to identify a model of the system, which is then used for model-based control, see \cite{Dorfler2022,Pasqualetti2021}. The main outcome of these works is that the indirect approach typically performs better if the offline data are only corrupted by measurement noise. In turn, direct data-based control can be superior if one faces a ``bias error", which can, e.g., result from identifying a linear model for a system which is, in fact, nonlinear, or when identifying a model of lower order than the real system dimension.

So far, these results are related to the control context. To the best of the authors' knowledge, no investigations regarding the dual problem of estimation are available in the literature. Therefore, we here show a first comparison between direct data-driven MHE and indirect data-driven MHE by means of numerical experiments. In Subsection~\ref{sec:Linear_System} we consider the linearized system with offline collected data that are subject to some additive measurement noise. In Subsection~\ref{sec:Nonlinear_system}, we consider exemplarily a bias error by identifying a four-dimensional linear model using data coming from a system that is, in fact, nonlinear.
\subsection{Linear System}
\label{sec:Linear_System}
The following linearized version of the open-loop stable four-tank system is considered
\begin{subequations}
	\label{example_system}
	\begin{align}
		x(t+1) =& 
		\begin{bmatrix}
			0.921 & 0 & 0.041 &0 \\
			0 & 0.918 & 0 & 0.033 \\
			0 & 0 & 0.924 & 0 \\
			0 & 0 & 0 & 0.937 \\
		\end{bmatrix}	x(t) \nonumber \\
		&+ \begin{bmatrix}
			0.017 & 0.001 \\
			0.001 & 0.023 \\
			0 & 0.061 \\
			0.072 & 0 \\
		\end{bmatrix}u(t) 
	\end{align}
	\begin{align}
		y(t)
		&= \begin{bmatrix}
			1 & 0 & 0 & 0 \\
			0 & 1 & 0 & 0 \\
		\end{bmatrix}
		x(t).
	\end{align}
	In the following, we comment on the design of the data-driven MHE. We start by collecting $N=100$ input, output and state measurements of system (\ref{example_system}), where both components of the persistently exciting input are sampled from a uniform random distribution $\mathcal{U}(0,1)$ (alternatively $\mathcal{U}(0,10)$).
\end{subequations}
\begin{figure}[t!]
	\centering
	\begin{minipage}{.5\textwidth}
		\centering
		\includegraphics[width=1\columnwidth]{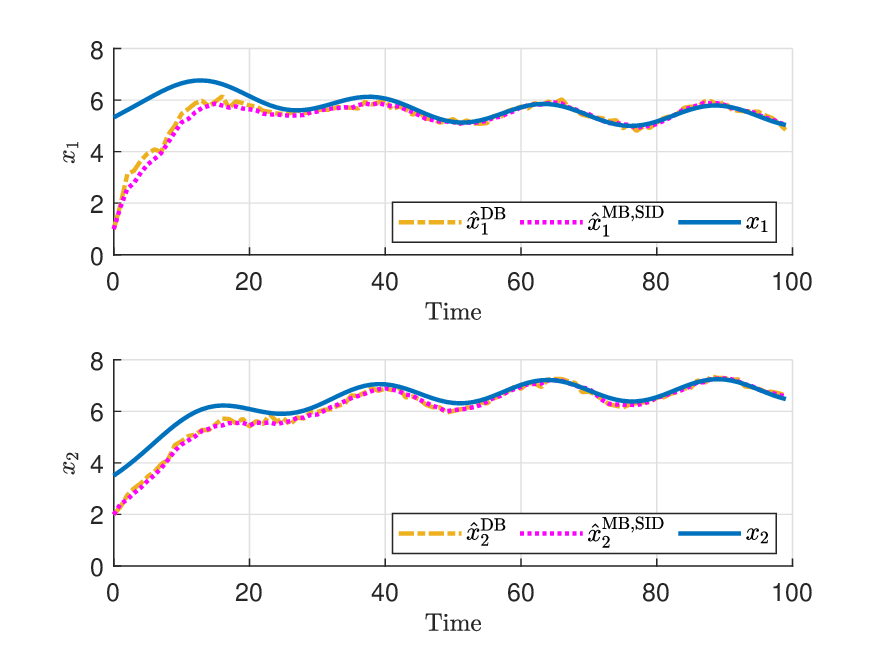}
	\end{minipage}
	\begin{minipage}{.5\textwidth}
		\centering
		\includegraphics[width=1\columnwidth]{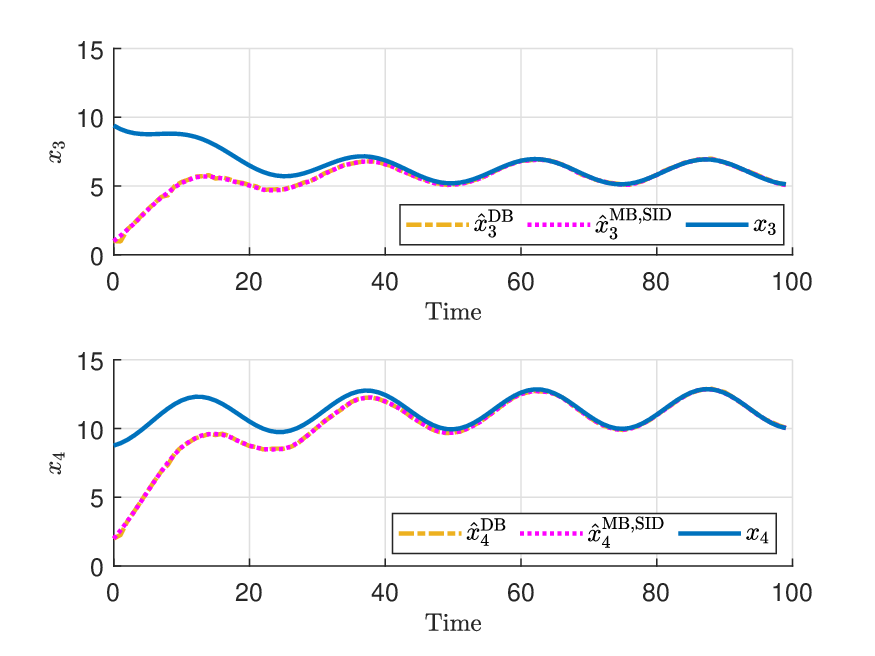}
	\end{minipage}
	\captionof{figure}{The simulation results illustrate the performance of the robust data-driven MHE scheme (where the estimated states are denoted by $\hat{x}^{\mathrm{DB}}$) defined in (\ref{MHE_noisy}) compared to a model-based MHE scheme (where the model is identified by SID techniques using the same data used in the data-driven MHE scheme and where the estimated states are denoted by $\hat{x}^{\mathrm{MB, SID}}$) for system (\ref{example_system}). The offline input is drawn from a uniform random distribution $\mathcal{U}(0,10)$ and $R$ is chosen as $R = 100$.}
	\label{fig:Simulation_Results}
\end{figure}
	The measurement noise regarding the offline collected outputs and states follows a truncated (at $\pm 0.003$) normal distribution with mean $\mu_{\mathcal{N}} = 0$ and standard deviation $\sigma = 0.001$. After the offline phase is completed, we construct the corresponding Hankel matrices using the measured inputs, states and outputs, and move to the online phase, where we apply a sinusoidal signal to both inputs.

	Regarding the cost function, we choose $\rho = 0.95$ (since $\eta = 0.9337$ for system (\ref{example_system}), where we determined the numerical value by means of the procedure outlined in Appendix~\ref{sec:db_eUOSS}, $c_\alpha = 2000$ and $c_{\sigma^x} = 600$. The upper bounds of the noise are chosen as $\overline{\varepsilon}_x^d = \overline{\varepsilon}_y^d =0.003$. The weighting matrices are set to $P  = 500 I_n$ and $R_1 = 100 I_p$ (alternatively $R_2 = 500 I_p$) and the state constraint set $\mathbb{X} = \{x \in \mathbb{R}^4 | x \geq 0\}$, i.e., the water levels in the four tanks cannot be negative. The noise~$v$ corrupting the online output measurements follows a normal distribution with $\mu_{\mathcal{N}} = 0$ and $\sigma = 0.5$. The horizon length (here $L=7$) is chosen such that the estimators achieve satisfactory performance in practice, while maintaining a reasonable computational complexity. 
		The prior estimate is chosen as $\hat{x}(0) = \begin{bmatrix}
			1 & 2 & 1& 2 \\
		\end{bmatrix}^\top$. 
		In Table~I, we show the results of~50 simulations per configuration (of~$R$ and the distribution from which $u^d$ is sampled). We consider 50 different online noise realizations and~50 different initial conditions (sampled from a uniform distribution over the interval $x(0)\in [0,10]^4$). For each noise realization and initial condition, the four different configurations are applied. Table~I shows the mean squared error (MSE) (averaged over the~50 simulations), defined as $\mathrm{MSE} \coloneqq \frac{1}{n T}\sum_{i=1}^{T} \sum_{j=1}^{n} (x_j(i)- \hat{x}_j(i))^2$, where $T$ denotes the number of time steps, here $T = 100$, and the mean absolute error (MAE) (also averaged over the 50 simulations), defined as $\mathrm{MAE} \coloneqq \frac{1}{n T}\sum_{i=1}^{T} \sum_{j=1}^{n} |x_j(i)- \hat{x}_j(i)|$ of all investigated combinations of the weighting matrix $R$ and distributions from which the offline control input is sampled. These results demonstrate that the robust data-driven MHE is pRES, as guaranteed by Theorem~\ref{thm:pRGES}. 
	\begin{table*}[t!]
		\label{tab:metrics_simulation}
		\centering
		\renewcommand{\arraystretch}{1.3}
		\begin{tabular}{llp{0.8cm}p{0.8cm}p{0.8cm}p{0.8cm}p{0.8cm}p{0.8cm}lp{0.8cm}p{0.8cm}p{0.8cm}p{0.8cm}p{0.8cm}p{0.8cm}} 
			\toprule
			& & \multicolumn{3}{c}{MSE} & \multicolumn{3}{c}{MAE} & & \multicolumn{3}{c}{MSE}& \multicolumn{3}{c}{MAE}\\  
			\cline{3-8}\cline{10-15}
			& & $\mathrm{MB}_{\mathrm{PEM}}$ & $\mathrm{MB}_{\mathrm{SID}}$ & $\mathrm{DD}$ &$\mathrm{MB}_{\mathrm{PEM}}$ & $\mathrm{MB}_{\mathrm{SID}}$ & $\mathrm{DD}$ & &$\mathrm{MB}_{\mathrm{PEM}}$ & $\mathrm{MB}_{\mathrm{SID}}$ & $\mathrm{DD}$ &$\mathrm{MB}_{\mathrm{PEM}}$ & $\mathrm{MB}_{\mathrm{SID}}$ & $\mathrm{DD}$ \\
			\cline{3-8}\cline{10-15}
			\parbox[t]{2mm}{\multirow{6}{*}{\rotatebox[origin=c]{90}{\textbf{Linear System}}}}& & \multicolumn{6}{c}{$u^d \sim \mathcal{U}(0,1)$}&\parbox[t]{2mm}{\multirow{6}{*}{\rotatebox[origin=c]{90}{\textbf{Nonlinear System}}}} & \multicolumn{6}{c}{$u^d \sim \mathcal{U}(0,1)$} \\
			&$R_1 = 100$ &  4.0388  & 1.0925 & \textbf{0.9909}  & 1.5909 & \textbf{0.4321} & 0.4780 & & \textbf{1.4716}  & 2.6269  & 2.5133 & \textbf{0.6810}  & 1.0933 & 0.9924   \\
			&$R_2 = 500 $ &  3.4661  & 0.8909  & \textbf{0.8292}  & 1.4714  & \textbf{0.3910}  & 0.4166 & & \textbf{1.3008}  & 2.4709  & 2.3364  & \textbf{0.6100}  & 1.0259  & 0.9567  \\
			\cline{3-8}\cline{10-15} 
			& & \multicolumn{6}{c}{$u^d \sim \mathcal{U}(0,10)$} & &  \multicolumn{6}{c}{$u^d \sim \mathcal{U}(0,10)$} \\
			&$R_1 = 100$ &  1.2065 & 1.1119 & \textbf{1.0676} & 0.5977 & 0.4365 & \textbf{0.4300} & & 1.7208  & 1.8795 & \textbf{1.6587} & 0.7150  & 0.7810 & \textbf{0.6938}   \\
			&$R_2 = 500 $ &  1.0938  & \textbf{0.9092}  & 0.9160 & 0.5751  & \textbf{0.3958} & 0.4112 & & 1.6012 & 1.5962 & \textbf{1.5268}  & 0.6974  & 0.6998 & \textbf{0.6541}  \\
			\bottomrule
		\end{tabular}
		\caption{Comparison of the performance of the proposed data-driven MHE scheme (denoted by DD) to model-based MHE schemes, where the model is identified using the same data by SID (denoted by $\mathrm{MB_{SID}}$) and PEM (denoted by $\mathrm{MB_{PEM}}$). We illustrate the performance for a nonlinear and linear example, two different uniform distributions from which the offline input is sampled, and two different choices of the weighting matrix $R$. We perform 50 simulations per configuration (with a different online noise realization and a different random initial condition) and show the averaged MSE and MAE over all simulations. Values in boldface denote the best values per configuration.}
	\end{table*}

	Regarding the design of the standard model-based MHE scheme, we first identify the matrices $A,~B,~C$ (applying \cite[Eq. (2.20)]{Van1997} to perform a SID and the built-in Matlab function $ssest$, which is based on PEM techniques, compare \cite{Ljung1998}) based on the same noisy data that we used to construct the Hankel matrices of the data-driven MHE scheme. Furthermore, we consider the same set-up, i.e., we chose the same weighing matrices $P$, $R_1$, $R_2$, the same discount factor~$\rho$, the same horizon length $L$, the same noise realizations, and the same initial conditions $x(0)$ and the same prior estimate~$\hat{x}(0)$.

	From the results illustrated in Table~I the linear system~(\ref{example_system}), we notice that the proposed data-driven MHE scheme has a similar performance to the model-based MHE scheme when the model is identified using SID. On the other hand, the performance of model-based MHE is worse when PEM-based identification is used, in particular for the case where $u^d$ is sampled from $\mathcal{U}(0,1)$. Depending on the selection of $u^d$ and $R$, either method can outperform the other. Hence, the data-driven MHE is a useful alternative that does not require model identification.
	
	Figure~\ref{fig:Simulation_Results} illustrates a typical behavior of the estimated states when the offline input data is sampled from $\mathcal{U}(0,10)$ and $R =100$. Due to space restrictions, we do not display figures for the other cases.

	\subsection{Nonlinear System}
	\label{sec:Nonlinear_system}
	From here on, we consider the original nonlinear system, introduced in \cite{Raff2006}
	\begin{subequations}
		\label{def_system_nonlinear}
		\begin{align}
			\dot{x}_1(t) & = -\frac{a_1}{A_1} \sqrt{2gx_1(t)} +\frac{a_3}{A_1} \sqrt{2gx_3(t)} + \frac{\gamma_1}{A_1} u_1(t) \\
			\dot{x}_2(t) & = -\frac{a_2}{A_2} \sqrt{2gx_2(t)} +\frac{a_4}{A_2} \sqrt{2gx_4(t)} + \frac{\gamma_2}{A_2} u_2(t) \\
			\dot{x}_3(t) & = -\frac{a_3}{A_3} \sqrt{2gx_3(t)} +\frac{1 - \gamma_2}{A_3} u_2(t) \\
			\dot{x}_4(t) & = -\frac{a_4}{A_4} \sqrt{2gx_4(t)} +\frac{1 - \gamma_1}{A_4} u_1(t).
		\end{align}
	\end{subequations}
	The numerical parameter values are given in \cite{Raff2006}. Obviously, for this nonlinear case the derived robust stability guarantees do not hold anymore.
	
	In the offline phase, we apply a persistently exciting input to the system which is sampled from a uniform distribution $\mathcal{U}(0,1)$ (alternatively $\mathcal{U}(0,10)$) and we collect $N= 100$ samples from system (\ref{def_system_nonlinear}). The measurement noise in the offline phase corrupting the output and state measurements follows a truncated (at $\pm 0.003$) normal distribution with $\mu_{\mathcal{N}} = 0$ and $\sigma = 0.001$. Once the data collection is completed, we switch to the online phase (in which we once again apply a sinusoidal signal). 
	
	The model-based and the data-driven MHE schemes are applied for the same parameters for the true system dynamics and the prior estimate, i.e., $\rho$, $\mathbb{X}$, $P$, $R_1$ (alternatively $R_2$), $L$, $x(0)$, $\hat{x}(0)$ remain the same, compared to Section \ref{sec:Linear_System}. Regarding the implementation of the data-driven MHE scheme, the numerical values of $c_\alpha$, $c_\sigma$, $\overline{ \varepsilon}_x$, $\overline{\varepsilon}_y$ remain also unchanged. In order to implement a standard model-based MHE scheme, we use the same noisy offline data collected from the nonlinear system~(\ref{def_system_nonlinear}) and identify a four-dimensional (discrete-time) linear system, by SID or PEM, such that we encounter a bias error. We simulate the different MHE schemes for 50 different online noise realizations and 50 different random initial conditions sampled from the interval $x(0) \in [0,10]^4$ as in the previous subsection.
	
The simulation results are illustrated in Table~I. In case~$u^d$ is sampled from a uniform distribution $\mathcal{U}(0,1)$, the model-based MHE scheme that is based on PEM yields the best performance, regardless of the selection of~$R$. However, the data-driven MHE scheme outperforms both model-based MHE schemes when~$u^d$ is sampled from a uniform distribution~$\mathcal{U}(0,10)$, i.e., in case of a larger excitation-to-noise ratio. In this example, exploring a larger region of the state-space in the offline phase improved the expressiveness of the data-driven system representation, such that the performance of the data-driven MHE scheme is better compared to the model-based MHE schemes. Thus, we again conclude that the proposed scheme is a competitive method that can outperform the alternatives.
	

\section{CONCLUSION}
\label{sec:Conclusion}
In this paper, we introduced a robust data-driven MHE framework that allows for noisy offline collected states and outputs, whose effect on the estimation is limited through regularization terms in the cost function. We showed pRES for the novel robust data-driven MHE scheme. A numerical example illustrated the performance of the robust data-driven MHE scheme. Furthermore, we compared its performance to a standard model-based MHE scheme, where the same noisy offline data was used.

Future work could focus on a variety of issues. First, our results are crucially based on the assumption that a potentially noisy state sequence is available from an offline phase. If one avoided this assumption, one could apply the robust MHE to a broader range of applications. Second, our theoretical guarantees are limited to linear systems, whereas MHE has been proven to be particular powerful for nonlinear systems. Therefore, extending the here presented results to (specific classes of) nonlinear systems would be interesting. Third, this paper did not focus on a theoretical comparison between data-driven and model-based MHE schemes. In the dual problem of MPC, some first results concerning a theoretical comparison of data- and model-based MPC were obtained in \cite{Dorfler2022,Pasqualetti2021}. An interesting topic for future research would be a similar analysis in the context of MHE.

\bibliographystyle{IEEEtran}
\bibliography{Literature_Robust_Data_Based_MHE}

\appendix

\subsection{Proof of Theorem~\ref{thm:pRGES}}
\label{proof}

\input{proof}

\input{computation_eUOSS_constants}

\end{document}

%% file: proof.tex
The framework of the proof is similar to robust stability proofs of various model-based MHE schemes, see \cite{Knufer2018,Allan2021robust,Mueller2017}. In turn, suitable adaptations to the data-driven setting are necessary such as, e.g., a representation of the system trajectories in a data-driven fashion.

\noindent \textbf{Part I: Establishment of a Contraction Mapping}

In this part of the proof, we consider $t < L$. The first step is once again to lower bound the optimal value function (exploiting Assumption~\ref{ass_stage}) for $t < L$ resulting in
\begin{align*}
	\allowdisplaybreaks
	J^\ast(\hat{x}(0|t),&\hat{\sigma}^y_{[-t, -1]}(t),\hat{\sigma}^x_{[-t, 0]}(t), \hat{\alpha}(t)) \nonumber \\
	\geq& p_0 |\hat{x}(0) - \hat{x}(0|t)|^2 \rho^t + r_0 \sum_{\tau = 1}^{t} |\hat{\sigma}^y(t-\tau|t)|^2 \rho^\tau \\ 
	&+ c_{\sigma^x}  |\hat{\sigma}^x_{[-t,0]}(t)|^2 +c_\alpha( \overline{\varepsilon}_x^d+\overline{\varepsilon}_y^d)|\hat{\alpha}(t)|^2.
\end{align*}
We want to apply Jensen's inequality \cite{Jensen1906}, which is defined as
\begin{align*}
	\varphi \left(\frac{\sum_i a_i x_i}{\sum_i a_i}\right) \leq \frac{\sum_i a_i \varphi(x_i)}{\sum_i a_i}
\end{align*}
for a convex function $\varphi$ and some positive $a_i$ and numbers $x_i$. In our case, the function $\varphi(y) \coloneqq y^2$. Hence, 
\begin{align*}
	\frac{(\sum_i a_i x_i)^2}{(\sum_i a_i)^2}= \left(\frac{\sum_i a_i x_i}{\sum_i a_i}\right)^2 \leq \frac{\sum_i a_i \varphi(x_i)}{\sum_i a_i}
\end{align*}
We multiply by $\sum_i a_i$ yielding
\begin{align*}
	\sum_i a_i \varphi(x_i) \geq \Big(\sum_i a_i\Big)^{-1} \Big(\sum_i a_i x_i\Big)^2
\end{align*}
We exploit this inequality to bound the optimal values functions and get
\begin{align}
	J^\ast(\hat{x}(0|t),&\hat{\sigma}^y_{[-t, -1]}(t),\hat{\sigma}^x_{[-t, 0]}(t), \hat{\alpha}(t)) \nonumber \\
	\geq& \Big(p_0\rho^t + r_0 \sum_{\tau = 1}^{t}\rho^\tau + c_{\sigma^x}+c_\alpha(\overline{\varepsilon}_x^d +\overline{\varepsilon}_y^d)\Big)^{-1} \times    \nonumber \\
	&\Big(p_0 |\hat{x}(0) - \hat{x}(0|t)| \rho^t + r_0 \sum_{\tau = 1}^{t} |\hat{\sigma}^y(t-\tau|t)| \rho^\tau \nonumber \\ 
	&+ c_{\sigma^x}  |\hat{\sigma}^x_{[-t,0]}(t)|  +c_\alpha (\overline{\varepsilon}_x^d +  \overline{\varepsilon}_y^d)|\hat{\alpha}(t)| \Big)^2 \label{eq:Jensen1_noisy}.
\end{align}
We define an auxiliary variable
\begin{align}
	c_{\tilde{J}} (t,\overline{\varepsilon}_x^d,&\overline{\varepsilon}_y^d) \coloneqq p_0\rho^t + r_0 \sum_{\tau =1}^{t}\rho^\tau + c_{\sigma^x}+ c_\alpha(\overline{\varepsilon}_x^d + \overline{\varepsilon}_y^d) \nonumber \\
	&= p_0 \rho^t + r_0 \frac{\rho - \rho^{t+1}}{1- \rho} +  c_{\sigma^x}+c_\alpha (\overline{\varepsilon}_x^d +  \overline{\varepsilon}_y^d). \label{def_cJ_tilde}
\end{align}
The introduced variable is now used to reformulate (\ref{eq:Jensen1_noisy})
\begin{align}
	(c_{\tilde{J}} &(t,\overline{\varepsilon}_x^d,\overline{\varepsilon}_y^d) J^\ast)^{1/2} \geq p_0 |\hat{x}(0) - \hat{x}(0|t)| \rho^t \nonumber \\
	&+ r_0  \sum_{\tau = 1}^{t} |\hat{\sigma}^y(t-\tau|t)| \rho^\tau  +  c_{\sigma^x}  |\hat{\sigma}^x_{[-t,0]}(t)|  \nonumber \\ 
	&+ c_\alpha (\overline{\varepsilon}_x^d +  \overline{\varepsilon}_y^d)|\hat{\alpha}(t)|. \label{eq:lower_final_noisy}
\end{align}
Next, we upper bound the optimal value function by means of a candidate trajectory, i.e., a feasible but in general suboptimal solution to problem (\ref{MHE_noisy}). Here we exploit the following candidate trajectory. We choose $\alpha(t)$ such that
\begin{equation}
	\begin{bmatrix}
		H_t(u^d_{[0, N-2]})\\
		H_1(x^d_{[0, N-t-1]})
	\end{bmatrix} \alpha(t) = 
	\begin{bmatrix}
		u_{[0, t-1]} \\
		x(0)
	\end{bmatrix},
	\label{alpha_candidate}
\end{equation}
where $x(0)$ denotes the real (unknown) system state $x$ at time $t=0$ and $u_{[0, t-1]} $ the real input sequence from time $0$ up to $t-1$. Please note that the left-hand side of (\ref{alpha_candidate}) contains the (unknown) \textit{noise-free} offline state sequence. According to Assumption~\ref{ass_pers} and Lemma~\ref{Lemma_Rank}, the concatenated Hankel matrices in equation (\ref{alpha_candidate}) have full row rank. Therefore, there exists a right inverse (denoted by $\dagger$) so that 
\begin{equation}
	\alpha(t) = \begin{bmatrix}
		H_t(u^d_{[0, N-2]})\\
		H_1(x^d_{[0, N-t-1]})
	\end{bmatrix}^\dagger
	\begin{bmatrix}
		u_{[0, t-1]} \\
		x(0)
	\end{bmatrix} \label{def_alpha}.
\end{equation}
The states and outputs of a DT-LTI system are linear combinations of the initial condition and the inputs (compare \cite{Markovsky2008} for a detailed explanation), therefore the above choice of $\alpha$ implies
\begin{equation}
	\begin{bmatrix}
		H_t(u^d_{[0, N-2]})\\
		H_t(y^d_{[0, N-2]})\\
		H_{t+1}(x^d_{[0, N-1]})
	\end{bmatrix} \alpha(t) = 
	\begin{bmatrix}
		u_{[0, t-1]} \\
		\tilde{y}_{[0, t-1]} - v_{[0, t-1]}\\
		x_{[0, t]} \\
	\end{bmatrix},
	\label{alpha_states_outputs}
\end{equation}
meaning that the real input, output, and state sequences are generated by that choice of $\alpha$, where we need to subtract the true measurement noise from the noisy output measurements in the second row of (\ref{alpha_states_outputs}) in order to make Willems' lemma hold. The slack $\sigma^x$ in relation to the candidate trajectory is chosen as
\begin{equation}
	\label{eq:sigmax}
	\sigma^x_{[-t, 0]}(t) = H_{t+1}(\varepsilon^d_{x,[0, N-1]}) \alpha(t) .
\end{equation}
The slack $\sigma^y$ must compensate the measurement noise (that occurs in the online phase) but also the noise affecting the offline collected output measurements. Therefore, the slack $\sigma^y$ associated with the candidate trajectory is chosen as
\begin{equation}
	\sigma^y_{[-t, -1]} (t) = \sigma^{\varepsilon_y}_{[-t, -1]} (t) + v_{[0, t-1]} \label{candidate_sigma_y}
\end{equation}
where $v$ corresponds to the true measurement noise and 
\begin{equation}
	\sigma^{\varepsilon_y}_{[-t, -1]} (t) = H_t(\varepsilon^d_{y, [0, N-2]}) \alpha(t). \label{eq:candidate_sigma_ey}
\end{equation}
This selection of $\alpha$, $\sigma^x$ and $\sigma^y$ implies that all constraints in (\ref{System_Dynamics_noisy}) are satisfied, meaning that the candidate trajectory is a feasible solution. Furthermore, (\ref{state_constraint_noisy}) holds since the real (undisturbed) state trajectory is contained in the constrained set $\mathbb{X}$. We want to upper bound the cost function related to the candidate solution. First, we can bound $|\alpha|$ by using (\ref{def_alpha}) to obtain
\begin{equation}
	|\alpha(t)| \leq \left|\begin{bmatrix}
		H_t(u^d_{[0, N-2]})\\
		H_1(x^d_{[0, N-t-1]})
	\end{bmatrix}^\dagger\right| \left|
	\begin{bmatrix}
		u_{[0, t-1]} \\
		x(0)
	\end{bmatrix} \right| .
\end{equation}
Since $u(t)$ and $x(t)$ evolve in a compact set (according to Assumption~\ref{ass_compact}), there exist a $u_\mathrm{max}$ and an $x_\mathrm{max}$ so that $  |u(t)| \leq u_\mathrm{max}$ and $ |x(t)| \leq x_\mathrm{max} \quad \forall t \in \mathbb{I}_{\geq 0}$. Additionally, we define
\begin{equation}
	H_{\mathrm{ux}} \coloneqq\left|\begin{bmatrix}
		H_t(u^d_{[0, N-2]})\\
		H_1(x^d_{[0, N-t-1]})
	\end{bmatrix}^\dagger\right|.
\end{equation}
In conclusion, it holds that
\begin{align}
	|\alpha(t)| &\leq H_{\mathrm{ux}}  (Lu_{\max} + x_{\max})\eqqcolon \alpha_{\mathrm{max}}. \label{final_alpha}
\end{align}
Hence, the variable $\alpha(t)$ associated with the candidate solution is upper bounded at any time $t$. Now, to upper bound the slack defined in (\ref{eq:sigmax}), we have
\begin{align}
	|\sigma^x_{[-t, 0]}(t)| &\leq |H_{t+1}(\varepsilon^d_{x,[0, N-1]})| |\alpha(t)|  \nonumber \\
	& \leq |H_{t+1}(\varepsilon^d_{x,[0, N-1]})|\alpha_{\mathrm{max}} \eqqcolon \sigma^x_\mathrm{max}. \label{final_sigma_x}
\end{align}
Finally, an upper bound for $\sigma^{\varepsilon_y}_{[-t, -1]}$ can be established by
\begin{align}
	|\sigma^{\varepsilon_y}_{[-t, -1]}(t)| &\leq |H_t(\varepsilon^d_{y, [0, N-2]}) \alpha(t)| \nonumber \\
	& \leq |H_t(\varepsilon^d_{y, [0, N-2]})| \alpha_{\max} \eqqcolon \sigma^{\varepsilon_y}_{\mathrm{max}} 
	\label{final_sigma_epsilon_y}
\end{align}
From here on, we consider the cost associated with the candidate trajectory defined by (\ref{alpha_candidate}), (\ref{eq:sigmax}), and (\ref{candidate_sigma_y}). This candidate trajectory can be upper bounded by using the basic inequality $(a+b)^2 \leq 2a^2 + 2b^2$ and Assumption~\ref{ass_stage} with $p_2 \coloneqq \lambda_{\mathrm{max}}(P)$ and $r_2 \coloneqq \lambda_{\mathrm{max}}(R)$  
\begin{align}
	J^\ast&(\hat{x}(0|t)) \leq J(x(0)) \leq p_2 |\hat{x}(0) - x(0)|^2 \rho^t \nonumber \\
	&+ 2r_2 \sum_{\tau = 1}^{t} |v(t-\tau)|^2 \rho^\tau + 2r_2 \sum_{\tau = 1}^{t} |\sigma^{\varepsilon_y}(t-\tau|t)|^2 \rho^\tau \nonumber\\ 
	&+  c_{\sigma^x}  |{\sigma}^x_{[-L,0]}(t)|^2  + c_\alpha (\overline{\varepsilon}_x^d  + \overline{\varepsilon}_y^d) |\alpha(t)|^2.
\end{align}
Note that we split the slack $\sigma^y$ into $v$ and $\sigma^{\varepsilon_y}$. In order to connect the established upper bound with the already derived lower bound, we need to multiply by $c_{\tilde{J}} (t,\overline{\varepsilon}_x^d,\overline{\varepsilon}_y^d)$ and to subsequently take the square root (using $\sqrt{a +b} \leq \sqrt{a} + \sqrt{b}$ which holds for any $a,b \geq 0$), yielding 
\begin{align}
	\big(c_{\tilde{J}} (t,\overline{\varepsilon}_x^d,&\overline{\varepsilon}_y^d) J^\ast(\hat{x}(0|t))\big)^{1/2} \nonumber \\
	\leq& \sqrt{ c_{\tilde{J}} (t,\overline{\varepsilon}_x^d,\overline{\varepsilon}_y^d)p_2} |\hat{x}(0) - x(0)| \rho^{t/2} \nonumber \\
	&+ \sqrt{ c_{\tilde{J}} (t,\overline{\varepsilon}_x^d,\overline{\varepsilon}_y^d)2r_2} \sum_{\tau = 1}^{t} |v(t-\tau)| \rho^{\tau/2} \nonumber\\ 
	&+ \sqrt{c_{\tilde{J}} (t,\overline{\varepsilon}_x^d,\overline{\varepsilon}_y^d)2r_2 }\sum_{\tau = 1}^{t} |\sigma^{\varepsilon_y}(t-\tau|t)|^2 \rho^{\tau /2} \nonumber\\ 
	&+ \sqrt{ c_{\tilde{J}} (t,\overline{\varepsilon}_x^d,\overline{\varepsilon}_y^d) c_{\sigma^x}}  |{\sigma}^x_{[-L,0]}(t)|  \nonumber\\ 
	&+ \sqrt{ c_{\tilde{J}}(t,\overline{\varepsilon}_x^d,\overline{\varepsilon}_y^d)c_\alpha (\overline{\varepsilon}_x^d  + \overline{\varepsilon}_y^d)} |\alpha(t)| \label{eq:upper_final_noisy} .
\end{align}
As in the previous section, we can combine the lower and the upper bound, namely
\begin{equation}
	\text {RHS of }(\ref{eq:lower_final_noisy})\leq \big(c_{\tilde{J}} (t,\overline{\varepsilon}_x^d,\overline{\varepsilon}_y^d)  J^\ast(\hat{x}(0|t))\big)^{1/2} \leq \text{RHS of }(\ref{eq:upper_final_noisy})
	\label{eq:final_cost_noisy},
\end{equation}
where the abbreviation RHS stands for ``right-hand side". 

The second key ingredient to prove pRES of the robust data-driven MHE is to exploit the e-UOSS property, which holds for system (\ref{system_def}) since it is assumed to be detectable (Assumption~\ref{ass_detec}), compare the discussion above Definition~\ref{Definition_UOSS_Text}. We here consider two trajectories. First, we consider the real (unknown) state trajectory parameterized by means of the candidate $\alpha$, compare (\ref{alpha_candidate}). This real state trajectory starts at $x_1 = x(0)$ and is driven by the input sequence $\mathbf{u}$. Its value at time $t$ corresponds to $x(t;x_1,\mathbf{u}) = x(t)$. Second, we consider a state trajectory based on the estimated state sequence. We cannot choose exactly the estimated state sequence (as done in the previous section) since this sequence is not necessarily a valid system trajectory due to the considered noise in the offline phase. We consider a sequence starting at 
\begin{align}
	x_2 &= \hat{x}(0|t)-H_1(\varepsilon^d_{x,{[0,N-t-1]}})\hat{\alpha}(t) + \hat{\sigma}^x(0|t) \nonumber \\ 
	&=H_1(x^d_{[0, N-t-1]})\hat{\alpha}(t),
\end{align}
which can be deduced from the last row of (\ref{System_Dynamics_noisy}). After $t$ time instances, the state value becomes
\begin{align}
	x(t;x_2,\mathbf{u}) &= \hat{x}(t) - H_1(\varepsilon_{x, [t, N-1]}^d)\hat{\alpha}(t) + \hat{\sigma}^x(t|t) \nonumber \\
	&=  H_1(x^d_{[t, N-1]}) \hat{\alpha}(t).
\end{align}
	Therefore,
	\begin{align}
		|x(t;x_1,\mathbf{u}) &-x(t;x_2,\mathbf{u})| = \nonumber \\
		& |x(t) - \hat{x}(t) + H_1(\varepsilon^d_{x,{[t,N-1]}})\hat{\alpha} -\hat{\sigma}^x(t|t)|.
	\end{align} 
	Note that $|a| - |b| \leq |a-b|$, which implies
	\begin{align*}
		|x(t) - \hat{x}(t)| &- |-H_1(\varepsilon_{x, [t, N-1]}^d)\hat{\alpha}(t) +\hat{\sigma}^x(t|t)| \\
		& \leq |x(t) - \hat{x}(t) + H_1(\varepsilon_{x, [t, N-1]}^d)\hat{\alpha}(t) -\hat{\sigma}^x(t|t)| \nonumber \\
		&= |x(t;x_1,\mathbf{u}) -x(t;x_2,\mathbf{u})|.
	\end{align*}
	Reformulating the above inequality yields
	\begin{align*}
		|x(t) - \hat{x}(t)| \leq &  |x(t;x_1,\mathbf{u}) -x(t;x_2,\mathbf{u})| \nonumber \\
		& + |-H_1(\varepsilon_{x, [t, N-1]}^d)\hat{\alpha}(t) +\hat{\sigma}^x(t|t)|. 
	\end{align*}
	Finally, we can replace the difference $|x(t;x_1,\mathbf{u}) -x(t;x_2,\mathbf{u})|$ by means of the e-UOSS definition (\ref{Definition_UOSS})
	\begin{align}
		|x&(t) - \hat{x}(t)| \nonumber \\ \leq& p_0 |x(0) -\hat{x}(0|t) + H_1(\varepsilon^d_{x, [0, N-t-1]})\hat{\alpha}(t) - \hat{\sigma}^x(0|t)| \eta^{t}  \nonumber \\
		&+ \sum_{\tau = 1}^{t} r_0|h(x(t-\tau; x_1,\mathbf{u}), u(t- \tau))  \nonumber \\
		&\quad \quad \quad- h(x(t-\tau; x_2,\mathbf{u}), u(t- \tau))| \eta^\tau \nonumber \\
		&+ |H_1(\varepsilon_{x, [t, N-1]}^d)\hat{\alpha}(t)| + |\hat{\sigma}^x(t|t)|. \label{eq:eUOSS_noisy1}
	\end{align}
	We can establish the following relationship for $0\leq t \leq L-1$, compare with the second row of (\ref{System_Dynamics_noisy})
	\begin{align}
		H_t(y_{[0, N-2]}^d)\hat{\alpha}(t) &+ H_t(\varepsilon_{y,[0, N-2]}^d)\hat{\alpha}(t) \nonumber \\ 
		&= \tilde{y}_{[0,t-1]} - \hat{\sigma}_{[-t, -1]}^y(t). \label{eq:output_eUOSS}
	\end{align}
	The expression for the nominal output trajectory (that is needed in the e-UOSS property) just corresponds to the first term of (\ref{eq:output_eUOSS}), namely $H_t(y_{[0, N-2]}^d)\hat{\alpha}(t)$. Therefore in the e-UOSS condition, we need to consider 
	\begin{align*}
		&\sum_{\tau = 1}^{t}h(x(t-\tau;x_2, \mathbf{u}), u(t-\tau)) =\sum_{\tau = 1}^{t} H_1(y_{[t-\tau,N -1-\tau]}) \hat{\alpha}(t) \\
		&=\sum_{\tau = 1}^{t}\big(  \tilde{y}(t-\tau) - \hat{\sigma}^y(t-\tau|t)- H_1(\varepsilon_{y,[t-\tau, N-1-\tau]}^d)\hat{\alpha}(t)\big) .
	\end{align*}
	Therefore, (\ref{eq:eUOSS_noisy1}) becomes
	\begin{align}
		|x&(t)- \hat{x}(t)| \nonumber \\
		\leq&   p_0 |x(0) - \hat{x}(0|t) + H_1(\varepsilon^d_{x, [0, N-t-1]})\hat{\alpha}(t) - \hat{\sigma}^x(0|t)| \eta^{t} \nonumber\\	
		&+ \sum_{\tau = 1}^{t} r_0 \Big |    
		\tilde{y}(t - \tau) - v(t-\tau)
		\nonumber \\ 
		&-\big(\tilde{y}(t-\tau) - \hat{\sigma}^y(t-\tau|t)  - H_1(\varepsilon^d_{y,[t-\tau, N-1-\tau]})\hat{\alpha}(t)   \big) \Big | \eta^\tau  \nonumber\\	
		&+ |H_1(\varepsilon_{x, [t, N-1]}^d)\hat{\alpha}(t)| + |\hat{\sigma}^x(t|t)| \nonumber\\[0.5em]
		\leq & p_0 |\hat{x}(0) - x(0)| \eta^{t} + p_0|\hat{x}(0|t)- \hat{x}(0)| \eta^{t} 
		\nonumber\\	
		&+ \sum_{\tau = 1}^{t} r_0|v(t-\tau)| \eta^\tau + \sum_{\tau = 1}^{t} r_0|\hat{\sigma}^y(t-\tau|t)| \eta^\tau \nonumber \\
		& + \sum_{\tau = 1}^{t}r_0 |H_1(\varepsilon^d_{y,[t-\tau, N-1-\tau]})\hat{\alpha}(t)| \eta^\tau 
		\nonumber\\	
		&+|H_1(\varepsilon_{x, [t, N-1]}^d)\hat{\alpha}(t)|+p_0|H_1(\varepsilon^d_{x, [0, N-t-1]})\hat{\alpha}(t)| \nonumber \\ 
		&+ |\hat{\sigma}^x(t|t)|  + p_0|\hat{\sigma}^x(0|t)|.  \label{e-UOSS_Basic_noisy2}
	\end{align}
	Next, the aim is to combine (\ref{e-UOSS_Basic_noisy2}) with the relation obtained in (\ref{eq:final_cost_noisy}). This is possible since 
	\begin{equation*}
		p_0|\hat{x}(0|t)- \hat{x}(0)| \eta^{t} \leq p_0 |\hat{x}(0) - \hat{x}(0|t)| \rho^t
	\end{equation*}
	and 
	\begin{equation*}
		r_0 \sum_{\tau = 1}^{t} |\hat{\sigma}^y(t-\tau|t)| \eta^\tau \leq r_0 \sum_{\tau = 1}^{t} r_0|\hat{\sigma}^y(t-\tau|t)|  \rho^\tau
	\end{equation*}
	by Assumption~\ref{ass_detec}, as well as 
	\begin{align*}
		|H_1(\varepsilon_{x, [t, N-1]}^d)\hat{\alpha}(t)| &\leq \sigma_{\max} (H_1(\varepsilon_{x, [t, N-1]}^d)) |\hat{\alpha}(t)| \nonumber \\ 
		& \leq || H_1(\varepsilon_{x, [t, N-1]}^d)||_F |\hat{\alpha}(t)| \nonumber \\
		&= \sqrt{\sum_{i=1}^{n} \sum_{j=1}^{N-t}  |\varepsilon^d_{x, i,j}|^2 }|\hat{\alpha}(t)| \nonumber \\
		&\leq \sqrt{n(N-t)} \overline{\varepsilon}_x^d |\hat{\alpha}(t)|\nonumber \\ 
		&\leq \frac{1}{2} c_\alpha \overline{\varepsilon}_x^d |\hat{\alpha}(t)|,
	\end{align*}
	where $\sigma_{\max}$ denotes the maximum singular value and $||\cdot||_F$ the Frobenius norm. 
	Similarly, we obtain
	\begin{align*}
		p_0|H_1(\varepsilon_{x, [0, N-t-1]}^d)\hat{\alpha}(t)| &\leq p_0 \sqrt{n(N-t)} \overline{\varepsilon}_x^d |\hat{\alpha}(t)| \nonumber \\ 
		&\leq \frac{1}{2} c_\alpha \overline{\varepsilon}_x^d |\hat{\alpha}(t)|.
	\end{align*}
	Furthermore, it holds that
	\begin{align*}
		\sum_{\tau = 1}^{t}&r_0 |H_1(\varepsilon^d_{y,[t-\tau, N-1-\tau]})\hat{\alpha}(t)| \eta^\tau  \nonumber \\
		& \leq \frac{\eta - \eta^{t+1}}{1-\eta}r_0\sqrt{p}  |H_t(\varepsilon^d_{y,[0, N-2]})|_\infty  |\hat{\alpha}(t)|  \nonumber \\
		& \leq \frac{\eta - \eta^{t+1}}{1-\eta}r_0 \sqrt{p}(N-1)\: \overline{\varepsilon}_y^d |\hat{\alpha}(t)|  \nonumber \\
		& \leq c_{\alpha}  \overline{\varepsilon}_y^d |\hat{\alpha}(t)|.
	\end{align*}
	Finally, we can upper bound the last terms of (\ref{e-UOSS_Basic_noisy2}) as
	\begin{equation*}
		|\hat{\sigma}^x(t|t)| + p_0 |\hat{\sigma}^x(0|t)|\leq c_{\sigma^x}  |\hat{\sigma}^x_{[-t,0]}(t)|
	\end{equation*}
	for $c_{\sigma^x}$ from (\ref{MHE_noisy_cost}). After combining the mentioned expressions, from (\ref{e-UOSS_Basic_noisy2}) we get
	\begin{align}
		|x&(t)- \hat{x}(t)|  \leq  p_0 |\hat{x}(0) -x(0)| \eta^{t} \nonumber \\
		& +p_0 |\hat{x}(0) - \hat{x}(0|t)| \rho^t + \sum_{\tau = 1}^{t} r_0|v(t-\tau)| \eta^\tau \nonumber \\
		&+  r_0 \sum_{\tau = 1}^{t} |\hat{\sigma}^y(t-\tau|t)|\rho^\tau \nonumber \\
		&+ c_{\alpha}  (\overline{\varepsilon}_y^d + \overline{\varepsilon}_x^d) |\hat{\alpha}(t)| + c_{\sigma^x}  |\hat{\sigma}^x_{[-t,0]}(t)|.
	\end{align}
	Now, we exploit the established relation (\ref{eq:final_cost_noisy}) to obtain 
	\begin{align}
		\allowdisplaybreaks
		|x&(t)- \hat{x}(t)|  \leq  p_0 |\hat{x}(0) -x(0)| \eta^{t}  + \sum_{\tau = 1}^{t} r_0|v(t-\tau)| \eta^\tau \nonumber \\
		&+\sqrt{ c_{\tilde{J}} (t,\overline{\varepsilon}_x^d,\overline{\varepsilon}_y^d)p_2} |\hat{x}(0) - x(0)| \rho^{t/2} \nonumber \\
		&+ \sqrt{ c_{\tilde{J}} (t,\overline{\varepsilon}_x^d,\overline{\varepsilon}_y^d)2r_2} \sum_{\tau = 1}^{t} |v(t-\tau)| \rho^{\tau/2} \nonumber \\
		&+ \sqrt{c_{\tilde{J}} (t,\overline{\varepsilon}_x^d,\overline{\varepsilon}_y^d)2r_2 }\sum_{\tau = 1}^{t} |\sigma^{\varepsilon_y}(t-\tau|t)| \rho^{\tau /2} \nonumber \\
		&+ \sqrt{ c_{\tilde{J}}(t,\overline{\varepsilon}_x^d,\overline{\varepsilon}_y^d)c_\alpha (\overline{\varepsilon}_x^d  + \overline{\varepsilon}_y^d)} |\alpha(t)| \nonumber \\
		&+ \sqrt{ c_{\tilde{J}} (t,\overline{\varepsilon}_x^d,\overline{\varepsilon}_y^d)c_{\sigma^x}}  |{\sigma}^x_{[-t,0]}(t)|
		\label{eq:comb_cost_e_UOSS_noisy}
	\end{align}
	We define 
	\begin{align*}
		&\tilde{P}_0 \coloneqq p_0 +\sqrt{c_{\tilde{J}} (t,\overline{\varepsilon}_x^d,\overline{\varepsilon}_y^d)p_2},  \\
		& \tilde{R}_0 \coloneqq r_0 +  \sqrt{ c_{\tilde{J}} (t,\overline{\varepsilon}_x^d,\overline{\varepsilon}_y^d)2r_2}, \\
		& \lambda \coloneqq \rho^{1/2},
	\end{align*}
	and simplify inequality (\ref{eq:comb_cost_e_UOSS_noisy}) as 
	\begin{align}
		|x&(t)- \hat{x}(t)|  \leq  \tilde{P}_0 |\hat{x}(0) -x(0)| \lambda^t  + \sum_{\tau = 1}^{t} \tilde{R}_0|v(t-\tau)| \eta^\tau \nonumber \\
		&+ \sqrt{c_{\tilde{J}} (t,\overline{\varepsilon}_x^d,\overline{\varepsilon}_y^d)2r_2 }\sum_{\tau = 1}^{t} |\sigma^{\varepsilon_y}(t-\tau|t)| \lambda^\tau \nonumber \\
		&+ \sqrt{ c_{\tilde{J}}(t,\overline{\varepsilon}_x^d,\overline{\varepsilon}_y^d)c_\alpha (\overline{\varepsilon}_x^d + \overline{\varepsilon}_y^d)} |\alpha(t)|  \nonumber \\
		&+ \sqrt{ c_{\tilde{J}} (t,\overline{\varepsilon}_x^d,\overline{\varepsilon}_y^d)c_{\sigma^x}}  |{\sigma}^x_{[-t,0]}(t)| .
		\label{eq:comb_cost_e_UOSS_noisy2}
	\end{align}
	Now, the objective is to bring (\ref{eq:comb_cost_e_UOSS_noisy2}) into the desired form (\ref{eq:pRGES}). To this end, we start by exploiting the upper bounds of the candidate trajectories in (\ref{final_alpha}), (\ref{final_sigma_x}), and (\ref{final_sigma_epsilon_y}), yielding
	\begin{align}
		|x&(t)- \hat{x}(t)|  \leq  \tilde{P}_0 |\hat{x}(0) -x(0)| \lambda^t  + \sum_{\tau = 1}^{t} \tilde{R}_0|v(t-\tau)| \eta^\tau \nonumber \\
		&+ \sqrt{c_{\tilde{J}} (t,\overline{\varepsilon}_x^d,\overline{\varepsilon}_y^d)2r_2 }\:\frac{\lambda-\lambda^{t+1}}{1- \lambda} \sigma^{\varepsilon_y}_{\mathrm{max}} \nonumber \\
		&+ \sqrt{ c_{\tilde{J}}(t,\overline{\varepsilon}_x^d,\overline{\varepsilon}_y^d)c_\alpha (\overline{\varepsilon}_x^d + \overline{\varepsilon}_y^d)} \: \alpha_{\mathrm{max}}  \nonumber \\
		&+ \sqrt{ c_{\tilde{J}} (t,\overline{\varepsilon}_x^d,\overline{\varepsilon}_y^d)c_{\sigma^x}} \sigma^x_\mathrm{max}.
		\label{eq:simple_after_candidate}
	\end{align}
	
	\noindent \textbf{Part II: Recursive Application of the Contraction Mapping}
	\indent From here on, we consider $t \geq L$. Therefore, $t$ must be replaced by $L$ in the definition of the candidate solution (compare (\ref{alpha_candidate}), (\ref{eq:sigmax}), and (\ref{candidate_sigma_y})) and in the definition of $c_{\tilde{J}}$ (see (\ref{def_cJ_tilde})). We reformulate (\ref{eq:simple_after_candidate}) for all $t\geq L$
	\begin{align}
		|x&(t)- \hat{x}(t)|  \leq  \tilde{P}_0 |\hat{x}(t-L) -x(t-L)| \lambda^L  \nonumber \\
		& + \sum_{\tau = 1}^{L} \tilde{R}_0|v(t-\tau)| \eta^\tau \nonumber \\
		&+ \sqrt{c_{\tilde{J}, \mathrm{max}}2r_2 }\: \frac{1-\lambda^{L+1}}{1- \lambda} \sigma^{\varepsilon_y}_{\mathrm{max}}\nonumber \\
		&+ \sqrt{c_{\tilde{J}, \mathrm{max}}c_\alpha (\overline{\varepsilon}_x^d + \overline{\varepsilon}_y^d)} \: \alpha_{\mathrm{max}}  \nonumber \\
		&+ \sqrt{ c_{\tilde{J}, \mathrm{max}}c_{\sigma^x}} \sigma^x_\mathrm{max},
		\label{eq:reformulation_noisy}
	\end{align}
	where $c_{\tilde{J}, \mathrm{max}} \coloneqq \max \{c_{\tilde{J}} (L,\overline{\varepsilon}_x^d,\overline{\varepsilon}_y^d),c_{\tilde{J}} (0,\overline{\varepsilon}_x^d,\overline{\varepsilon}_y^d)\}$. Inequality (\ref{eq:reformulation_noisy}) establishes a contraction mapping from $t-L$ up to time $t$ for $L$ large enough. The remaining steps regarding the terms $x - \hat{x}$ and $v$ are completely analogous to the previous section. Therefore, these steps are not explained in detail here. We focus on the additional terms that appear due to the considered noise in the offline phase. We choose $L$ large enough so that \begin{align}
		\tilde{\mu} \coloneqq \tilde{P}_0 \lambda^L \in (0,1).
	\end{align}
	We apply the established contraction mapping recursively for $t = TL + \tilde{t}$
	\begin{align}
		|x(t) &- \hat{x}(t)| \leq \tilde{\mu}^T | x(t- TL) - \hat{x}(t- TL)| \nonumber \\ &
		+ R_0 \sum_{k = 0}^{T-1} \tilde{\mu}^k \sum_{\tau = 1}^{L}  |v(t - kL-\tau)| \lambda^\tau \nonumber \\ & + \sum_{k=0}^{T-1} \tilde{\mu}^k(c_4(\overline{\varepsilon}_x^d, \overline{\varepsilon}_y^d) \frac{\lambda-\lambda^{L+1}}{1- \lambda} +c_5(\overline{\varepsilon}_x^d, \overline{\varepsilon}_y^d))  \label{eq:noisy_nearly_final}
	\end{align}
	for 
	\begin{align}
		c_4(\overline{\varepsilon}_x^d, \overline{\varepsilon}_y^d)   \coloneqq& \sqrt{c_{\tilde{J}, \mathrm{max}}2r_2 }\:  \sigma^{\varepsilon_y}_{\mathrm{max}} \label{def_c4}  \\
		c_5(\overline{\varepsilon}_x^d, \overline{\varepsilon}_y^d)\coloneqq&  \sqrt{ c_{\tilde{J}, \mathrm{max}}c_\alpha (\overline{\varepsilon}_x^d  + \overline{\varepsilon}_y^d)} \: \alpha_{\mathrm{max}}  \nonumber \\
		&+ \sqrt{c_{\tilde{J}, \mathrm{max}}c_{\sigma^x}} \sigma^x_\mathrm{max}. \label{def_c5}
	\end{align}
	We again replace the first term of inequality (\ref{eq:noisy_nearly_final}) with the expression obtained in (\ref{eq:simple_after_candidate}) and get
	\begin{align*}
		|x(t) &- \hat{x}(t)| \leq \tilde{\mu}^T \Big(\tilde{P}_0 |\hat{x}(0) -x(0)| \lambda^{\tilde{t}}  + \sum_{\tau = 1}^{\tilde{t}} \tilde{R}_0|v(\tilde{t}-\tau)| \eta^\tau \nonumber \\
		&+ c_4(\overline{\varepsilon}_x^d, \overline{\varepsilon}_y^d)\frac{\lambda -\lambda^{\tilde{t}+1}}{1- \lambda} + c_5(\overline{\varepsilon}_x^d, \overline{\varepsilon}_y^d)\Big) 	\nonumber \\ 
		& + R_0 \sum_{k = 0}^{T-1} \tilde{\mu}^k \sum_{\tau = 1}^{L}  |v(t - kL-\tau)| \lambda^\tau \nonumber \\ & 
		+ \sum_{k=0}^{T-1} \tilde{\mu}^k(c_4(\overline{\varepsilon}_x^d, \overline{\varepsilon}_y^d) \frac{ \lambda -\lambda^{L+1}}{1- \lambda} +c_5(\overline{\varepsilon}_x^d, \overline{\varepsilon}_y^d)) .
	\end{align*}
	The terms related to $|\hat{x}(0) - x(0)|$ and $v$ can be reformulated as in the previous section. We define $\check{\lambda} \coloneqq \tilde{\mu}^{1/L}$ and obtain
	\begin{flalign*}
		\allowdisplaybreaks
		|x(t)- \hat{x}(t)|  \leq&  \tilde{P}_0 |\hat{x}(0) -x(0)| \check{\lambda}^t   + \sum_{\tau = 1}^{t} \tilde{R}_0|v(t-\tau)| \check{\lambda}^\tau \nonumber \\
		& + \tilde{\mu}^T (c_4(\overline{\varepsilon}_x^d, \overline{\varepsilon}_y^d)\frac{\lambda -\lambda^{\tilde{t}+1}}{1- \lambda} + c_5(\overline{\varepsilon}_x, \overline{\varepsilon}_y^d)) \nonumber\\
		&+ \sum_{k=0}^{T-1}\tilde{\mu}^k(c_4(\overline{\varepsilon}_x^d, \overline{\varepsilon}_y^d) \frac{\lambda-\lambda^{L+1}}{1- \lambda} +c_5(\overline{\varepsilon}_x^d, \overline{\varepsilon}_y^d)) . 
	\end{flalign*}
	Since $\tilde{t} < L$, this expression can be upper bounded by applying the geometric series
	\begin{flalign*}
		|x(t)- &\hat{x}(t)|  \leq\tilde{P}_0 |\hat{x}(0) -x(0)| \check{\lambda}^t   + \sum_{\tau = 1}^{t} \tilde{R}_0|v(t-\tau)| \check{\lambda}^\tau \nonumber \\
		& + \frac{1 - \tilde{\mu}^{T+1}}{1 - \tilde{\mu}} (c_4(\overline{\varepsilon}_x^d, \overline{\varepsilon}_y^d) \frac{\lambda-\lambda^{L+1}}{1- \lambda} +c_5(\overline{\varepsilon}_x^d, \overline{\varepsilon}_y^d)).  
	\end{flalign*}
	Finally, we define
	\begin{align}
		\gamma (\overline{\varepsilon}^d) \coloneqq  \frac{1}{1 - \tilde{\mu}} (c_4(\overline{\varepsilon}^d, \overline{\varepsilon}^d) \frac{\lambda-\lambda^{L+1}}{1- \lambda} +c_5(\overline{\varepsilon}^d, \overline{\varepsilon}^d)),
		\label{def_gamma}
	\end{align}
	such that $\gamma \in \mathcal{K}$. In particular, it holds that $\gamma(0) = 0$, since $c_4(0,0) = c_5(0,0) = 0$. This can been seen from (\ref{def_c4}) and (\ref{def_c5}). For $\overline{\varepsilon}^d = 0$, it follows that $\sigma^{\varepsilon_y}_{\mathrm{max}} = 0$, and $\sigma^x_{\mathrm{max}} = 0$ (compare eqs. (\ref{final_sigma_epsilon_y}) and (\ref{final_sigma_x})), as well as $\overline{\varepsilon}_x^d =\overline{\varepsilon}_y^d = 0$. Note that
	\begin{align*}
		\frac{1- \tilde{\mu}^{T+1}}{1 - \tilde{\mu}} (c_4(\overline{\varepsilon}_x^d, \overline{\varepsilon}_y^d) \frac{\lambda-\lambda^{L+1}}{1- \lambda}+c_5(\overline{\varepsilon}_x^d, \overline{\varepsilon}_y^d)) \leq \gamma (\overline{\varepsilon}^d).
	\end{align*}
	We obtain 
	\begin{flalign}
		|x(t)- \hat{x}(t)|  \leq&  \tilde{P}_0 |\hat{x}(0) -x(0)| \check{\lambda}^t   + \sum_{\tau = 1}^{t} \tilde{R}_0|v(t-\tau)| \check{\lambda}^\tau \nonumber \\
		& +\gamma (\overline{\varepsilon}^d), \label{final_error_bounds}
	\end{flalign}
	which proves that the data-driven MHE is pRES. \hfill $\blacksquare$

%% file: computation_eUOSS_constants.tex
\subsection{Data-Driven Computation of e-UOSS Constants}
	\label{sec:db_eUOSS}
	Based on \cite[Corollary 7]{Knufer2020}, one can deduce that $\eta$ can be chosen as $\eta = |A_L|_{P_e}$, where $A_L$ corresponds to $A_L = A+LC$ (with $L$ such that $A +LC$ is Schur stable) and $P_e$ to the solution of \cite[Eq. (28)]{Knufer2020}. Furthermore, the constants $p_0$ and $r_0$ can be chosen as $p_0 = |P_e^{1/2}|/\sqrt{\lambda_{\mathrm{min}}(P_e)}$ and $r_0 = |L|_{P_e} |P_e^{1/2}|/(\lambda_{\mathrm{min}}(P_e) |A_L|_{P_e})$. To obtain a data-driven expression of $A_L$ (which then allows to compute $P_e$ by means of \cite[Eq. (28)]{Knufer2020}), we design a data-driven state feedback controller\footnote{Note that the controller design requires data of the original system for which $\mathrm{rank}\Bigg(\begin{bmatrix} H_1(x^d_{[1,N-1]}) \\ H_1(y^d_{[0,N-2]}) \end{bmatrix}\Bigg) = n + p$ holds.} $L^\top$ for the dual system, as explained in \cite{Adachi2021,De2019}. This controller leads to stable closed-loop dynamics, i.e., $A^\top + C^\top L^\top$ is Schur stable. Then, we take advantage of the duality between control and estimation. In fact, the stabilizing controller for the dual system is at the same time an observer for the original system with stable error dynamics. Furthermore, the closed-loop dynamics of the stabilized dual system (represented by $A^\top + C^\top L^\top$ in the model-based system representation) correspond to $A+LC$ of the original system, which allows to determine the required e-UOSS constants. Please note that this procedure is valid for noise-free (offline) data. Nevertheless, in simulations, we observe that this procedure still produces Schur stable matrices $A_L$ (enabling a data-driven computation of the e-UOSS constants) as long as the noise level is small enough.

%% file: arxive_revised.bbl
\begin{thebibliography}{10}
\providecommand{\url}[1]{#1}
\csname url@samestyle\endcsname
\providecommand{\newblock}{\relax}
\providecommand{\bibinfo}[2]{#2}
\providecommand{\BIBentrySTDinterwordspacing}{\spaceskip=0pt\relax}
\providecommand{\BIBentryALTinterwordstretchfactor}{4}
\providecommand{\BIBentryALTinterwordspacing}{\spaceskip=\fontdimen2\font plus
\BIBentryALTinterwordstretchfactor\fontdimen3\font minus
  \fontdimen4\font\relax}
\providecommand{\BIBforeignlanguage}[2]{{%
\expandafter\ifx\csname l@#1\endcsname\relax
\typeout{** WARNING: IEEEtran.bst: No hyphenation pattern has been}%
\typeout{** loaded for the language `#1'. Using the pattern for}%
\typeout{** the default language instead.}%
\else
\language=\csname l@#1\endcsname
\fi
#2}}
\providecommand{\BIBdecl}{\relax}
\BIBdecl

\bibitem{Kalman1960}
R.~E. Kalman, ``A new approach to linear filtering and prediction problems,''
  \emph{Transactions of the ASME–Journal of Basic Engineering}, vol.~82, pp.
  35--45., 1960.

\bibitem{Luenberger1971}
D.~Luenberger, ``An introduction to observers,'' \emph{IEEE Transactions on
  Automatic Control}, vol.~16, no.~6, pp. 596--602, 1971.

\bibitem{Rawlings2020}
J.~B. Rawlings, D.~Q. Mayne, and M.~Diehl, \emph{Model predictive control:
  theory, computation, and design}, 2nd~ed.\hskip 1em plus 0.5em minus
  0.4em\relax Nob Hill Publishing Madison, WI, 2020.

\bibitem{Rao2001}
C.~V. Rao, J.~B. Rawlings, and J.~H. Lee, ``Constrained linear state
  estimation—a moving horizon approach,'' \emph{Automatica}, vol.~37, no.~10,
  pp. 1619--1628, 2001.

\bibitem{Alessandri2014}
A.~Alessandri and M.~Awawdeh, ``Moving-horizon estimation for discrete-time
  linear systems with measurements subject to outliers,'' in \emph{53rd IEEE
  Conference on Decision and Control}, 2014, pp. 2591--2596.

\bibitem{Gharbi2019}
M.~Gharbi and C.~Ebenbauer, ``Proximity moving horizon estimation for linear
  time-varying systems and a bayesian filtering view,'' in \emph{2019 IEEE 58th
  Conference on Decision and Control (CDC)}.\hskip 1em plus 0.5em minus
  0.4em\relax IEEE, 2019, pp. 3208--3213.

\bibitem{Farina2010}
M.~Farina, G.~Ferrari-Trecate, and R.~Scattolini, ``Distributed moving horizon
  estimation for linear constrained systems,'' \emph{IEEE Transactions on
  Automatic Control}, vol.~55, no.~11, pp. 2462--2475, 2010.

\bibitem{Willems2005}
J.~C. Willems, P.~Rapisarda, I.~Markovsky, and B.~L. De~Moor, ``A note on
  persistency of excitation,'' \emph{Systems \& Control Letters}, vol.~54,
  no.~4, pp. 325--329, 2005.

\bibitem{Markovsky2008}
I.~Markovsky and P.~Rapisarda, ``Data-driven simulation and control,''
  \emph{International Journal of Control}, vol.~81, no.~12, pp. 1946--1959,
  2008.

\bibitem{De2019}
C.~De~Persis and P.~Tesi, ``Formulas for data-driven control: Stabilization,
  optimality, and robustness,'' \emph{IEEE Transactions on Automatic Control},
  vol.~65, no.~3, pp. 909--924, 2019.

\bibitem{Coulson2019}
J.~Coulson, J.~Lygeros, and F.~D{\"o}rfler, ``Data-enabled predictive control:
  In the shallows of the {DeePC},'' in \emph{2019 18th European Control
  Conference (ECC)}.\hskip 1em plus 0.5em minus 0.4em\relax IEEE, 2019, pp.
  307--312.

\bibitem{Berberich2020_MPC}
J.~Berberich, J.~K{\"o}hler, M.~A. M{\"u}ller, and F.~Allg{\"o}wer,
  ``Data-driven model predictive control with stability and robustness
  guarantees,'' \emph{IEEE Transactions on Automatic Control}, vol.~66, no.~4,
  pp. 1702--1717, 2020.

\bibitem{Markovsky2021}
I.~Markovsky and F.~D{\"o}rfler, ``Behavioral systems theory in data-driven
  analysis, signal processing, and control,'' \emph{Annual Reviews in Control},
  vol.~52, pp. 42--64, 2021.

\bibitem{Lopez2022}
V.~G. Lopez and M.~A. M{\"u}ller, ``On a continuous-time version of willems’
  lemma,'' in \emph{2022 IEEE 61st Conference on Decision and Control
  (CDC)}.\hskip 1em plus 0.5em minus 0.4em\relax IEEE, 2022, pp. 2759--2764.

\bibitem{Turan2021}
M.~S. Turan and G.~Ferrari-Trecate, ``Data-driven unknown-input observers and
  state estimation,'' \emph{IEEE Control Systems Letters}, vol.~6, pp.
  1424--1429, 2022.

\bibitem{Shi2022}
J.~Shi, Y.~Lian, and C.~N. Jones, ``Data-driven input reconstruction and
  experimental validation,'' \emph{IEEE Control Systems Letters}, vol.~6, pp.
  3259--3264, 2022.

\bibitem{Adachi2021}
R.~Adachi and Y.~Wakasa, ``Dual system representation and prediction method for
  data-driven estimation,'' in \emph{2021 60th Annual Conference of the Society
  of Instrument and Control Engineers of Japan (SICE)}.\hskip 1em plus 0.5em
  minus 0.4em\relax IEEE, 2021, pp. 1245--1250.

\bibitem{Alanwar2021}
A.~Alanwar, A.~Berndt, K.~H. Johansson, and H.~Sandberg, ``Data-driven
  set-based estimation using matrix zonotopes with set containment
  guarantees,'' in \emph{2022 European Control Conference (ECC)}.\hskip 1em
  plus 0.5em minus 0.4em\relax IEEE, 2022, pp. 875--881.

\bibitem{liu2023learning}
W.~Liu, J.~Sun, G.~Wang, F.~Bullo, and J.~Chen, ``Learning robust data-based
  {LQG} controllers from noisy data,'' \emph{arXiv preprint arXiv:2305.01417},
  2023.

\bibitem{Wolff2021}
T.~M. Wolff, V.~G. Lopez, and M.~A. M{\"u}ller, ``Data-based moving horizon
  estimation for linear discrete-time systems,'' in \emph{2022 European Control
  Conference (ECC)}.\hskip 1em plus 0.5em minus 0.4em\relax IEEE, 2022, pp.
  1778--1783.

\bibitem{Berberich2020_trajectory}
J.~Berberich and F.~Allg{\"o}wer, ``A trajectory-based framework for
  data-driven system analysis and control,'' in \emph{2020 European Control
  Conference (ECC)}.\hskip 1em plus 0.5em minus 0.4em\relax IEEE, 2020, pp.
  1365--1370.

\bibitem{VanWarde2020}
H.~J. van Waarde, C.~De~Persis, M.~K. Camlibel, and P.~Tesi, ``Willems’
  fundamental lemma for state-space systems and its extension to multiple
  datasets,'' \emph{IEEE Control Systems Letters}, vol.~4, no.~3, pp. 602--607,
  2020.

\bibitem{Graeber2019}
T.~Gräber, S.~Lupberger, M.~Unterreiner, and D.~Schramm, ``A hybrid approach
  to side-slip angle estimation with recurrent neural networks and kinematic
  vehicle models,'' \emph{IEEE Transactions on Intelligent Vehicles}, vol.~4,
  no.~1, pp. 39--47, 2019.

\bibitem{Ehlers2022}
S.~F.~G. Ehlers, Z.~Ziaukas, J.-P. Kobler, and H.-G. Jacob, ``State and
  parameter estimation in a semitrailer for different loading conditions only
  based on trailer signals,'' in \emph{2022 American Control Conference (ACC)},
  2022, pp. 2353--2360.

\bibitem{Van1997}
P.~Van~Overschee and B.~De~Moor, \emph{Subspace identification for linear
  systems: Theory—Implementation—Applications}.\hskip 1em plus 0.5em minus
  0.4em\relax Kluwer Academic Publishers, 1997.

\bibitem{Mueller2017}
M.~A. M{\"u}ller, ``Nonlinear moving horizon estimation in the presence of
  bounded disturbances,'' \emph{Automatica}, vol.~79, pp. 306--314, 2017.

\bibitem{Allan2019}
D.~A. Allan and J.~B. Rawlings, ``Moving horizon estimation,'' in
  \emph{Handbook of Model Predictive Control}, S.~V. Rakovi{\'{c}} and W.~S.
  Levine, Eds.\hskip 1em plus 0.5em minus 0.4em\relax Cham: Springer
  International Publishing, 2019, pp. 99--124.

\bibitem{Knufer2018}
S.~Knüfer and M.~A. Müller, ``Robust global exponential stability for moving
  horizon estimation,'' in \emph{2018 IEEE Conference on Decision and Control
  (CDC)}, 2018, pp. 3477--3482.

\bibitem{Knuefer2021}
S.~Kn{\"u}fer and M.~A. M{\"u}ller, ``Nonlinear full information and moving
  horizon estimation: Robust global asymptotic stability,'' \emph{Automatica},
  vol. 150, p. 110603, 2023.

\bibitem{Schiller2022suboptimal}
J.~D. Schiller and M.~A. M{\"u}ller, ``Suboptimal nonlinear moving horizon
  estimation,'' \emph{IEEE Transactions on Automatic Control}, vol.~68, no.~4,
  pp. 2199--2214, 2022.

\bibitem{Coulson2019b}
J.~Coulson, J.~Lygeros, and F.~D{\"o}rfler, ``Regularized and distributionally
  robust data-enabled predictive control,'' in \emph{2019 IEEE 58th Conference
  on Decision and Control (CDC)}.\hskip 1em plus 0.5em minus 0.4em\relax IEEE,
  2019, pp. 2696--2701.

\bibitem{Knufer2020}
S.~Kn{\"u}fer and M.~A. M{\"u}ller, ``Time-discounted incremental
  input/output-to-state stability,'' in \emph{2020 59th IEEE Conference on
  Decision and Control (CDC)}.\hskip 1em plus 0.5em minus 0.4em\relax IEEE,
  2020, pp. 5394--5400.

\bibitem{Allan2021robust}
D.~A. Allan and J.~B. Rawlings, ``Robust stability of full information
  estimation,'' \emph{SIAM Journal on Control and Optimization}, vol.~59,
  no.~5, pp. 3472--3497, 2021.

\bibitem{Alessandri2012}
A.~Alessandri, M.~Baglietto, and G.~Battistelli, ``Min-max moving-horizon
  estimation for uncertain discrete-time linear systems,'' \emph{SIAM Journal
  on Control and Optimization}, vol.~50, no.~3, pp. 1439--1465, 2012.

\bibitem{Schiller2022moving}
J.~D. Schiller and M.~A. M{\"u}ller, ``A moving horizon state and parameter
  estimation scheme with guaranteed robust convergence,'' \emph{arXiv preprint
  arXiv:2211.09053}, 2022.

\bibitem{Pan2021}
G.~Pan, R.~Ou, and T.~Faulwasser, ``On a stochastic fundamental lemma and its
  use for data-driven optimal control,'' \emph{IEEE Transactions on Automatic
  Control}, 2022.

\bibitem{Raff2006}
T.~Raff, S.~Huber, Z.~K. Nagy, and F.~Allgower, ``Nonlinear model predictive
  control of a four tank system: An experimental stability study,'' in
  \emph{2006 IEEE International Conference on Control Applications}.\hskip 1em
  plus 0.5em minus 0.4em\relax IEEE, 2006, pp. 237--242.

\bibitem{Dorfler2022}
F.~D{\"o}rfler, J.~Coulson, and I.~Markovsky, ``Bridging direct \& indirect
  data-driven control formulations via regularizations and relaxations,''
  \emph{IEEE Transactions on Automatic Control}, 2022.

\bibitem{Pasqualetti2021}
V.~Krishnan and F.~Pasqualetti, ``On direct vs indirect data-driven predictive
  control,'' in \emph{2021 60th IEEE Conference on Decision and Control (CDC)},
  2021, pp. 736--741.

\bibitem{Ljung1998}
L.~Ljung, ``System identification,'' in \emph{Signal analysis and
  prediction}.\hskip 1em plus 0.5em minus 0.4em\relax Springer, 1998, pp.
  163--173.

\bibitem{Jensen1906}
J.~L. W.~V. Jensen, ``Sur les fonctions convexes et les in{\'e}galit{\'e}s
  entre les valeurs moyennes,'' \emph{Acta mathematica}, vol.~30, no.~1, pp.
  175--193, 1906.

\end{thebibliography}
